\documentclass[numberedappendix]{emulateapj}

\usepackage{graphicx}
\usepackage{epsfig}
\usepackage{amssymb}
\usepackage{epsf}
\usepackage{natbib}

\newcommand{\vecalp}{\vec{\alpha}}
\newcommand{\veceta}{\vec{\eta}}

\newcommand{\vecx}{\vec{x}}
\newcommand{\vecy}{\vec{y}}
\newcommand{\vecb}{\vec{b}}
\newcommand{\vecn}{\vec{n}}
\newcommand{\vecxi}{\vec{\xi}}

\newcommand{\veclam}{\vec{\lambda}}

\newcommand{\matC}{\mathbf{C}}

\newcommand{\matA}{\mathbf{A}}
\newcommand{\matAt}{\mathbf{A}^{\rm T}}
\newcommand{\matB}{\mathbf{B}}

\newcommand{\matHx}{\mathbf{H}_{x}}
\newcommand{\matHxt}{\mathbf{H}_{x}^{\rm T}}
\newcommand{\matHy}{\mathbf{H}_{y}}
\newcommand{\matHyt}{\mathbf{H}_{y}^{\rm T}}
\newcommand{\matKx}{\mathbf{K}_{L}}
\newcommand{\matKxt}{\mathbf{K}_{L}^{\rm T}}
\newcommand{\matKy}{\mathbf{K}_{E}}
\newcommand{\matKyt}{\mathbf{K}_{E}^{\rm T}}

\newcommand{\matR}{\mathbf{R}}

\newcommand{\slos}{\sigma^{2}_{\rm los}}
\newcommand{\penf}{\mathcal{P}}

\newcommand{\vc}{v_{\rm c}}
\newcommand{\Rcore}{R_{\rm s}}
\newcommand{\Rc}{R_{\rm c}}

\newcommand{\model}{\mathcal{M}}

\newcommand{\Ed}{E_{\mathcal{L}}}
\newcommand{\Es}{E_{\mathcal{R}}}
\newcommand{\Ep}{E_{\mathcal{P}}}
\newcommand{\Nd}{N_{\rm d}}

\newcommand{\Ns}{N_{\rm s}}
\newcommand{\Ng}{N_{\gamma}}
\newcommand{\Nb}{N_{\rm b}}
\newcommand{\Nx}{N_{\rm x}}
\newcommand{\nTIC}{N_{\rm TIC}}
\newcommand{\nE}{N_{E}}
\newcommand{\nLz}{N_{L_{z}}}
\newcommand{\nSB}{N_{\SB}}
\newcommand{\nvz}{N_{\vz}}
\newcommand{\mTIC}{M_{\rm TIC}}

\newcommand{\Cv}{\matC^{-1}}
\newcommand{\CvL}{\matC^{-1}_{\rm L}}
\newcommand{\CvD}{\matC^{-1}_{\rm D}}

\newcommand{\Cj}{\mathcal{C}_{j}}
\newcommand{\de}{{\rm d}}
\newcommand{\PA}{\vartheta_{\rm PA}}
\newcommand{\vz}{\langle v_{z'} \rangle}
\newcommand{\vphi}{\langle v_{\varphi} \rangle}
\newcommand{\vqz}{\langle v^{2}_{z'} \rangle}
\newcommand{\vqzz}{\langle v^{2}_{z} \rangle}
\newcommand{\vqR}{\langle v^{2}_{R} \rangle}
\newcommand{\vqphi}{\langle v^{2}_{\varphi} \rangle}
\newcommand{\dynlen}{\textsc{cauldron}}
\newcommand{\evid}{\mathcal{E}}
\newcommand{\Phiz}{\Phi_{0}}
\newcommand{\tPhi}{\tilde{\Phi}}

\newcommand{\tz}{\tilde{z}}
\newcommand{\tx}{\tilde{x}'}
\newcommand{\ty}{\tilde{y}'}
\newcommand{\txi}{\tilde{\vecxi}}
\newcommand{\talp}{\alpha_{0}}

\newcommand{\Lz}{L_{z}}
\newcommand{\DF}{f(E, \Lz)}
\newcommand{\zd}{z_{\rm d}}
\newcommand{\zs}{z_{\rm s}}
\newcommand{\Dd}{D_{\rm d}}
\newcommand{\Ds}{D_{\rm s}}
\newcommand{\Dds}{D_{\rm ds}}
\newcommand{\SB}{\Sigma}
\newcommand{\Azvc}{A_{{\rm ZVC}, j}}
\newcommand{\AELz}{A_{[E_{j}, L_{z,j}]}}

\newcommand{\machine}{{3 Ghz}}
\newcommand{\Nrealiz}{{100}}

\newcommand{\vecd}{\vec{d}}
\newcommand{\mockd}{\vec{d}_{\rm sim}}

\newcommand{\vecp}{\vec{p}}
\newcommand{\mockp}{\vec{p}_{\rm sim}}

\newcommand{\vecs}{\vec{s}}
\newcommand{\mocks}{\vec{s}_{\rm sim}}
\newcommand{\recs}{\vec{s}_{\rm rec}}
\newcommand{\vecgam}{\vec{\gamma}}
\newcommand{\mockgam}{\vec{\gamma}_{\rm sim}}
\newcommand{\recgam}{\vec{\gamma}_{\rm rec}}

\newcommand{\matL}{\mathbf{L}}
\newcommand{\matLt}{\mathbf{L}^{\rm T}}
\newcommand{\matM}{\mathbf{M}}
\newcommand{\matMt}{\mathbf{M}^{\rm T}}
\newcommand{\matQ}{\mathbf{Q}}
\newcommand{\matQt}{\mathbf{Q}^{\rm T}}

\newcommand{\matH}{\mathbf{H}}
\newcommand{\matHt}{\mathbf{H}^{\rm T}}

\newcommand{\lamlen}{\lambda_{\rm len}}

\newcommand{\lamx}{\lambda^{\rm dyn}_{L}}
\newcommand{\lamy}{\lambda^{\rm dyn}_{E}}

\shorttitle{Combining Gravitational Lensing and Stellar Dynamics}
\shortauthors{M. Barnab\`e \& L. V. E. Koopmans}

\begin{document}

\title{A unifying framework for self-consistent gravitational lensing\\ 
  and stellar dynamics analyses of early-type galaxies}

\author{Matteo Barnab\`e and L\'eon V. E. Koopmans}

\affil{Kapteyn Astronomical Institute, University of Groningen, P.O. Box 800, 
       9700 AV Groningen, The Netherlands}

\begin{abstract}
Gravitational lensing and stellar dynamics are two independent
methods, based solely on gravity, to study the mass distributions of
galaxies. Both methods suffer from degeneracies, however, that are
difficult to break. In this paper, we present a new framework that
self-consistently unifies gravitational lensing and stellar dynamics,
breaking some classical degeneracies that have limited their
individual usage, particularly in the study of high-redshift galaxies.

For any given galaxy potential, the mapping of both the unknown lensed
source brightness distribution and the stellar phase-space
distribution function onto the photometric and kinematic observables
can be cast as a single set of coupled linear equations, which are
solved by maximizing the likelihood penalty function. The Bayesian
evidence penalty function subsequently allows one to find the best
potential-model parameters and to quantitatively rank potential-model
families or other model assumptions (e.g. PSF). We have implemented a
fast algorithm that solves for the maximum-likelihood pixelized lensed
source brightness distribution and the two-integral stellar
phase-space distribution function $f(E,L_z)$, assuming axisymmetric
potentials. To make the method practical, we have devised a new
Monte Carlo approach to Schwarzschild's orbital superposition method,
based on the superposition of two-integral ($E$ and $L_z$) toroidal
components, to find the maximum-likelihood two-integral distribution
function in a matter of seconds in any axisymmetric potential. The
non linear parameters of the potential are subsequently found through
a hybrid MCMC and Simplex optimization of the evidence. Illustrated by
the power-law potential models of Evans, we show that the inclusion of
stellar kinematic constraints allows the correct linear and non linear
model parameters to be recovered, including the potential strength,
oblateness {\sl and} inclination, which, in the case of
gravitational-lensing constraints only, would otherwise be fully
degenerate.
\end{abstract}

\keywords{gravitational lensing --- stellar dynamics --- galaxies:
structure --- galaxies: elliptical and lenticular, cD}

\section{Introduction}
\label{intro}

Understanding the formation and the evolution of early-type galaxies
is one of the most important open problems in present-day astrophysics
and cosmology.  Within the standard $\Lambda$CDM paradigm, massive
ellipticals are predicted to be formed via hierarchical merging of
lower mass galaxies \citep[][]{Toom.77, Frenk.88, White.91, Barnes.92,
Cole.00}. Despite the many theoretical and observational successes of
this scenario, several important features of early-type galaxies are
still left unexplained. In particular, the origin of the (often
strikingly tight) empirical scaling laws that correlate the global
properties of ellipticals remain unexplained: (i) the fundamental
plane \citep{Djor.87, Dres.87}, relating effective radius, velocity
dispersion and effective surface brightness; (ii) the $M_{\rm
BH}-\sigma$ \citep{Mago.98, Ferr.00, Gebh.00}, relating the mass of
the central supermassive black hole hosted by the galaxy with its
velocity dispersion; (iii) the color-$\sigma$ \citep*{Bower.92} and
the ${\rm Mg}_{2} - \sigma$ \citep*{Guzm.92, Bend.93, Bern.03}
relating the velocity dispersion with the stellar ages and the
chemical composition of the galaxy. 

Each of these scaling relations relate structural,
(spectro-) photometric, and dynamical (i.e.\ stellar velocity dispersion)
quantities. Whereas the first two are solely based on the observed
stellar component, the latter is a function of the stellar {\sl and}
dark matter mass distribution. Hence, a detailed study of the inner
mass profile of early-type galaxies at different redshifts is
undoubtedly necessary to properly address the numerous issues related
with the formation and the evolution of these objects and their scaling
relations, and it would also constitute an excellent test bed for the
validity of the $\Lambda$CDM scenario on small (i.e.\ non linear)
scales.

Today, a large number of thorough stellar dynamic and X-ray studies
have been conducted to probe the mass structure of nearby ($z \la
0.1$) early-type galaxies \citep*{Fabb.89, Mould.90, Sagl.92, Bert.94,
Franx.94, Caro.95, Arna.96, Rix.97, Mats.98, Loew.99, Gerh.01,
Selj.02, deZe.02, Borr.03, Roma.03, Capp.06, Gava.07}, in most (but
not all) cases finding evidence for the presence of a dark matter halo
component and for a flat equivalent rotation curve in the inner
regions.

When it comes to distant ($z \ga 0.1$) early-type galaxies, however,
only relatively little is known. The two main diagnostic tools which
can be employed to carry out such studies, namely gravitational
lensing \citep[][]{saasfee} and stellar dynamics
\citep[][]{Binn.book}, both suffer from limitations and
degeneracies. Gravitational lensing provides an accurate and almost
model independent determination of the total mass projected within the
Einstein radius, but a reliable determination of the mass density
profile is often prevented by the effects of the mass sheet
\citep*{Falco.85} and the related mass profile and shape degeneracies
\citep[e.g.][]{Wuck.02, Evans.03, Saha.06}. Despite these limitations,
strong gravitational lensing has provided the first, although
heterogeneous, constraints on the internal mass distribution of stellar
and dark matter in high-redshift galaxies \citep[e.g.][]{Koch.95,
Rusin.01, Ma.03, Rusin.02, Cohn.01, Munoz.01, Winn.03, Wuck.04,
Fer.05, Wayth.05, Dye.05, Brewer.06b, Dobke.06}.

Analyses based on stellar dynamics are limited by the paucity of
bright kinematic tracers in the outer regions of early-type galaxies
\citep[e.g.][]{Gerh.06}.  Moreover, there is a severe degeneracy (the
mass-anisotropy degeneracy) between the mass profile of the galaxy and
the anisotropy of its stellar velocity dispersion tensor
\citep[e.g.][]{Gerh.93}.  Higher order velocity moments, which could
be used to overcome this difficulty, cannot be measured in reasonable
integration time for such distant systems with any of the current
instruments.

A viable solution to break these degeneracies to a large extent is a
joint analysis which combines constraints from both gravitational
lensing and stellar dynamics \citep{Koop.02, Treu.02, Treu.03,
Koop.03, Rusin.03, Treu.04, Rusin.05, Bolton.06, Treu.06, Koop.06,
Gava.07}: the two approaches can be described as almost
``orthogonal'', in other words they complement each other very
effectively \citep[see e.g.][for some simple scaling
relations]{Koop.04}, allowing a trustworthy determination of the mass
density profile of the lens galaxy in the inner regions (i.e.\ within
the Einstein radius or the effective radius, whichever is larger). A
joint gravitational lensing and stellar dynamics study of the Lenses
Structure and Dynamics (LSD) and Sloan Lens ACS (SLACS) samples of
early-type galaxies \citep[][]{Koop.06, Bolton.06, Treu.06}, for
example, showed that the total mass density has an average logarithmic
slope extremely close to isothermal ($\langle s' \rangle = 2.01\pm
0.03$, assuming $\rho_{\rm tot} \propto r^{-s'}$) with a very small
intrinsic spread of at most 6\%. They also find no evidence for any
secular evolution of the inner density slope of early-type galaxies
between $z = 0.08$ and $1.01$ \citep[][]{Koop.06}.

These authors, in their combined analysis, make use of both
gravitational lensing and stellar dynamics information, but treat
the two problems as {\sl independent}. More specifically, the
projected mass distribution of the lens galaxy is modeled as a
singular isothermal ellipsoid \citep[SIE; e.g.][]{Korm.94}, and the
obtained value for the total mass within the Einstein radius is used
as a constraint for the dynamical modeling; in the latter the galaxy
is assumed to be spherical and the stellar orbit anisotropy to be
constant \citep[or follow an Osipkov--Merritt profile;][]{Osip.79,
Merr.85a, Merr.85b} in the inner regions and therefore the spherical
Jeans equations can be solved in order to determine the slope $s'$
\citep[see][for a full discussion of the methodology]{Koop.03, Treu.04,
Koop.06}.  This approach, however, while being robust and successful,
is not fully self-consistent: gravitational lensing and the stellar
dynamics use different assumptions about the symmetry of the system
and different potentials for the lens galaxy.

This has motivated us to develop a completely rigorous and
self-consistent framework to carry out combined gravitational lensing
and stellar dynamics studies of early-type galaxies, the topic of the
present paper. The methodology that we introduce, in principle, is
completely general and allows -- given a set of data from
gravitational lensing (i.e.\ the surface brightness distribution of
the lensed images) and stellar dynamics (i.e.\ the surface brightness
distribution and the line-of-sight projected velocity moments of the
lens galaxy) -- the ``best'' parameters, describing the gravitational
potential of the galaxy and its stellar phase-space distribution
function, to be recovered.

In practice, because of technical and computational limitations, we
restrict ourselves to axisymmetric potentials and two-integral
stellar phase-space distribution functions \citep[i.e.\ $f(E,L_z)$;
see][]{Binn.book}, in order to present a fast and efficiently working
algorithm. We introduce a new Monte Carlo approach to Schwarzschild's
orbital superposition method, that allows the $f(E,L_z)$ to be solved
in axisymmetric potentials in a matter of seconds. All of these
restrictions, however, should be seen just as one particular
implementation of the much broader general framework.

The very core of the methodology lies in the possibility of
formulating both the lensed image reconstruction and the dynamical
modeling as formally analogous linear problems. The whole framework is
coherently integrated in the context of Bayesian statistics, which
provides an objective and unambiguous criterion for model comparison,
based on the \emph{evidence} merit function
\citep{MacKay.92,MacKay.03}. The Bayesian evidence penalty function
allows one to both determine the ``best'' (i.e. the most plausible in
an Occam's razor sense) set of non linear parameters for an assigned
potential and compare and rank different potential families, as well
as set the optimal regularization level in solving the linear
equations and find the best point-spread function (PSF) model, pixel
scales, etc.

The paper is organized as follows: In Sect.~\ref{bayes} we first
review some relevant aspects of the theory of Bayesian inference, and
we elucidate how these apply to the framework of combined lensing and
stellar dynamics. In Sect.~\ref{outline} we present a general overview
of the framework, with particular focus on the case of our
implementation for axisymmetric potentials. In Sects.~\ref{len}
and~\ref{dyn}, respectively, we provide a detailed description of the
methods for the lensed image reconstruction and the dynamical
modeling.  In Sect.~\ref{test} we describe the testing of the method,
showing that it allows a reliable recovery of the correct model
potential parameters. Finally, conclusions are drawn in
Sect.~\ref{conc}.  The application of the algorithm to the early-type
lens galaxies of the SLACS sample \citep[e.g.][]{Bolton.06, Treu.06,
Koop.06, Gava.07} will be presented in forthcoming papers.

\section{Data fitting and model comparison:\\ A Bayesian approach} 
\label{bayes}

Before describing the code itself in more detail, in this Section
we first review the basics of Bayesian statistics. This is relevant,
because -- as will be shown in more detail in the coming Sections --
both the lensing and stellar dynamics parts of the algorithm can be
formalized as linear sets of equations
\begin{equation}
\label{Ax+n=b}
\vecb = \matA \vecx + \vecn 
\end{equation}
that can be solved within a Bayesian statistical framework. In the
above equation, $\vecb$ represents the data, $\matA =
\matA[\Phi(\veceta)]$ is the model, which will in general depend on
the physics involved (e.g., in our case, the lens-galaxy potential
$\Phi$, a function of the non linear parameters $\veceta$), $\vecx$
are the linear parameters that we want to infer, and $\vecn$ is the
noise in the data, characterized by the covariance matrix $\matC$.

First, we aim to determine the linear parameters $\vecx$ given the
data and a fixed model and, on a more general level, to find the
non linear parameters $\veceta$ corresponding to the ``best'' model.
Note that the choices of grid size, PSF, etc., are also regarded as
being (discrete) parameters of the model family (changing these
quantities or assumptions formally is equivalent to adopting a
different family of models).

Both of these problems can quantitatively be addressed within the
framework of Bayesian statistics. Following the approach of MacKay
(\citeauthor{MacKay.92} \citeyear{MacKay.92}, \citeyear{MacKay.99},
\citeauthor{MacKay.03} \citeyear{MacKay.03}; see also
\citeauthor{Suyu.06} \citeyear{Suyu.06}), it is possible to
distinguish between three different levels of inference for the data
modeling.

\begin{enumerate}

\item At the first level of inference, the model $\matA$ is assumed to
be true, and the most probable $\vecx_{\rm MP}$ for the linear
parameters is obtained by maximizing the posterior probability, given
by the expression (derived from Bayes' rule)
\begin{equation}
\label{bayes.1}
P(\vecx | \vecb, \lambda, \matA, \matH) = 
\frac{P(\vecb | \vecx, \matA) P(\vecx | \matH, \lambda)}
     {P(\vecb | \lambda, \matA, \matH)} ,
\end{equation}
where $\matH$ is a regularization operator, which formalizes a
conscious a priori assumption about the degree of smoothness that we
expect to find in the solution (refer to Appendix~\ref{reg} for the
construction of the regularization matrix); the level of
regularization is set by the hyperparameter value $\lambda$. The
introduction of some kind of prior (the term $P(\vecx | \matH,
\lambda)$ in Eq.~[\ref{bayes.1}]) is inescapable, since, due to the
presence of noise in the data, the problem from
equation~(\ref{Ax+n=b}) is an ill-conditioned linear system and,
therefore, cannot simply be solved through a direct inversion along
the line of $\vecx = \matA^{-1} \vecb$ (see for example
\citeauthor{NR.92} \citeyear{NR.92} for a clear introductory treatment
on this subject).  Note that besides $\matH$, any a priori assumption
(PSF, pixel scales, etc.) can be treated and ranked through the
evidence. Traditional likelihood methods do not allow a proper
quantitative ranking of model families or other assumptions. In
Eq.~(\ref{bayes.1}), the probability $P(\vecb | \vecx, \matA)$ is the
likelihood term, while the normalization constant $P(\vecb | \lambda,
\matA, \matH)$ is called the evidence and plays a fundamental role in
Bayesian analysis, because it represents the likelihood term at the
higher levels of inference.

\item At the second level, we infer the most probable hyperparameter
$\lambda_{\rm MP}$ for the model $\matA$ by maximizing the posterior
probability function
\begin{equation}
\label{bayes.2}
P(\lambda | \vecb, \matA, \matH) = 
\frac{P(\vecb | \lambda, \matA, \matH) P(\lambda)}
     {P(\vecb | \matA, \matH)} ,
\end{equation}
which is equivalent to maximizing the likelihood $P(\vecb | \lambda,
\matA, \matH)$ (i.e.\ the evidence of the previous level) if the prior
$P(\lambda)$ is taken to be flat in logarithm, as is customarily
done,\footnote{With the statistics terminology this is an
``uninformative'' prior, which tries to represent the absence of a
priori information on the scale of $\lambda$ (see
e.g. \citeauthor{Cous.95} \citeyear{Cous.95} and references
therein).}, because its scale is not known.

\item Finally, at the third level of inference the models are objectively
compared and ranked on the basis of the evidence (of the previous
level),
\begin{equation}
\label{bayes.3}
P(\matA, \matH | \vecb) \propto 
P(\vecb | \matA, \matH) P(\matA, \matH),
\end{equation}
where the prior $P(\matA, \matH)$ is assumed to be flat. It has been
shown by \citet{MacKay.92} that this evidence-based Bayesian method
for model comparison automatically embodies the principle of Occam's
razor, i.e.\ it penalizes those models which correctly predict the
data, but are unnecessarily complex. Hence, it is in some sense
analogous to the reduced $\chi^{2}$.

\end{enumerate}

In Sections~\ref{lin.opt} and~{\ref{nlin.opt}} we illustrate how this
general framework is implemented under reasonable simplifying
assumptions and how the maximization of the evidence is done in
practice.

\subsection{Maximizing the posterior: Linear optimization}
\label{lin.opt}

If the noise $\vecn$ in the data can be modeled as Gaussian, it is
possible to show \citep{MacKay.92, Suyu.06} that the posterior
probability (Eq.~[\ref{bayes.1}]) can be written as
\begin{equation}
\label{posterior}
P(\vecx | \vecb, \lambda, \matA, \matH) = 
\frac{\exp \left[ -\Ep(\vecx)  \right]}
{\int \exp \left[ -\Ep(\vecx)  \right] \de \vecx} ,
\end{equation}
where 
\begin{equation}
\label{pen.f}
\Ep = \Ed(\vecx) + \lambda \Es(\vecx) .
\end{equation}

In the penalty function from equation~(\ref{pen.f}),
\begin{equation}
\label{likelihood}
\Ed(\vecx) = 
\frac{1}{2} {(\matA \vecx - \vecb)}^{\rm T} \Cv (\matA \vecx - \vecb) 
\end{equation}
(i.e.\ half of the $\chi^{2}$ value) is a term proportional to the
logarithm of the likelihood, which quantifies the agreement of the
model with the data, and the term
\begin{equation}
\label{reg.func}
\Es(\vecx) = \frac{1}{2} {|| \matH \vecx ||}^{2}
\end{equation}
is the regularizing function, which is taken to have a quadratic form
with the minimum in $\vecx_{\rm reg} = \vec{0}$ (see the seminal paper
of \citeauthor{Tikh.63}~\citeyear{Tikh.63} for the use of the
regularization method in solving ill-posed problems). The term
$\Es$ formalizes the only a priori assumption regarding the smoothness
of the solution, and therefore corresponds to the prior term of
Eq.~(\ref{bayes.1}).

The most probable solution $\vecx_{\rm MP}$ is obtained by maximizing
the posterior from equation~(\ref{posterior}). The calculation of
$\partial \left[ \Ed(\vecx) + \lambda \Es(\vecx)\right] / \partial
\vecx = \vec{0}$ yields the linear set of normal equations
\begin{equation}
\label{ATA.2}
(\matAt \Cv \matA + \lambda \matHt \matH) \vecx = \matAt \Cv \vecb ,
\end{equation}
%
which maximizes the posterior (a solution exists and is unique because
of the quadratic and positive definite nature of the matrices).

If the solution $\vecx_{\rm MP}$ is unconstrained, the set of
equations~(\ref{ATA.2}) can be effectively and non-iteratively solved
using e.g.\ a Cholesky decomposition technique. However, in our case
the solutions have a very direct physical interpretation, representing
the surface brightness distribution of the source in the case of
lensing (e.g.\ Section~\ref{len}), or the weighted distribution function
in the case of dynamics (e.g.\ Section~\ref{dyn}). The solutions
must therefore be non-negative. Hence, we compute the solution of the
constrained set of equations~(\ref{ATA.2}) using the freely-available
L-BFGS-B code, a limited memory and bound constrained implementation
of the Broyden-Fletcher-Goldfarb-Shanno (BFGS) method for solving
optimization problems \citep*{Byrd.95, Zhu.97}.

\begin{figure*}[!t]
\begin{center}
\resizebox{0.90\hsize}{!}{\includegraphics{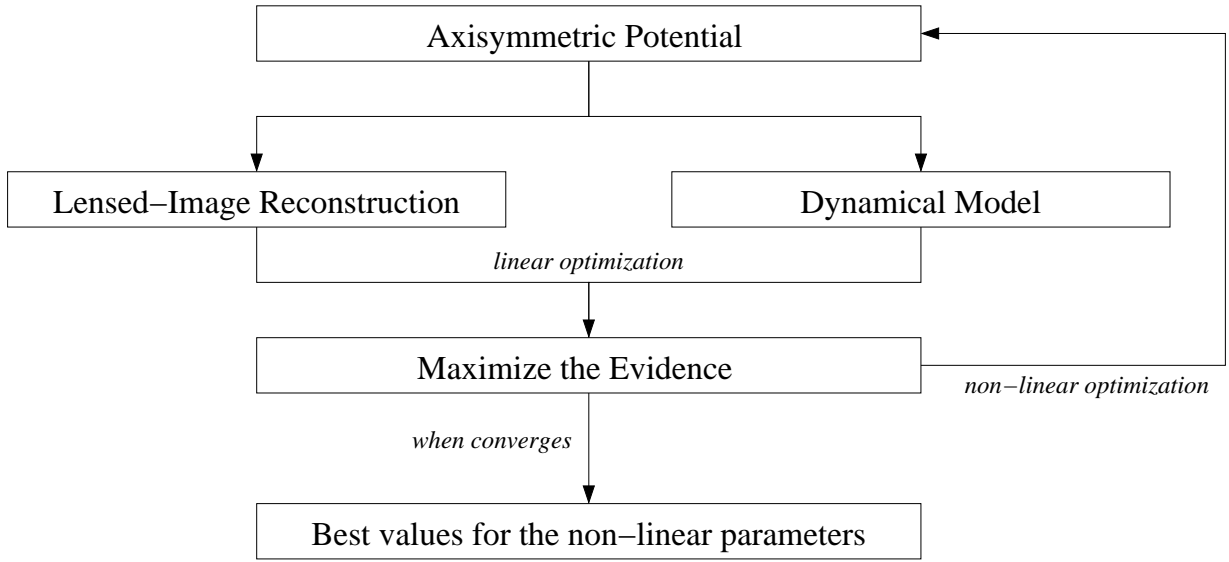}}
\caption{Scheme of the general framework for joint gravitational
lensing and stellar dynamics analysis. See the text for an extended
description.}
\label{scheme}
\end{center}
\end{figure*}

\subsection{Maximizing the evidence: non linear optimization}
\label{nlin.opt}

The implementation of the higher levels of inference (i.e.\ the model
comparison) requires an iterative optimization process. At every
iteration~$i$, a different set $\veceta_{i}$ of the non linear
parameters is considered, generating a new model $\matA_{i} \equiv
\matA[\Phi(\veceta_{i})]$ for which the most probable solution
$\vecx_{{\rm MP},i}$ for the linear parameters is recovered as
described in Sect.~\ref{lin.opt}, and the associated evidence
$\evid(\veceta_{i}, \lambda) = P(\vecb | \lambda, \matA_{i}, \matH)$
is calculated (the explicit expression for the evidence is
straightforward to compute but rather cumbersome and is, therefore,
given in Appendix~\ref{evidence}). Then a nested loop, corresponding
to the second level of inference, is carried out to determine the most
probable hyperparameter $\lambda_{\rm MP}$ for this model, by
maximization of $\evid(\veceta_{i}, \lambda)$. The evidence $\evid_{i}
= P(\vecb|\matA_{i}, \matH)$ for the model $\matA_{i}$ can now be
calculated by marginalizing over the
hyperparameters.\footnote{However, it turns out that a reasonable
approximation is to calculate $\evid_{i}$ assuming that $P(\lambda |
\vecb, \matA_{i}, \matH)$ is a $\delta$ function centered on
$\lambda_{\rm MP}$, so that the evidence is obtained simply as $
P(\vecb | \lambda_{\rm MP}, \matA_{i}, \matH)$
\citep[see][]{MacKay.92, Suyu.06}.} The different models $\matA_{i}$
can now be ranked according to the respective value of the evidence
$\evid_{i}$ (this is the third level of inference), and the best model
is the one which maximizes the evidence. This procedure is in fact
very general, and can be applied to compare models with different
types of potentials, regularization matrices, PSFs, grid sizes, etc.

\subsubsection{The Hyperparameters}

In principle, for each $\veceta_{i}$ one has to determine the
corresponding set of most probable hyperparameters $\lambda_{{\rm MP},
i}$ by means of a nested optimization loop.  In practice, however, the
values of the hyperparameters change only slightly when $\veceta$ is
varied in the optimization process, and therefore it is not necessary
to perform a nested loop for $\lambda$ at each $\veceta$-iteration.
What we do is to start the evidence maximization by iteratively
changing $\veceta$, while the hyperparameters are kept fixed at a
quite large initial value $\lambda_{0}$ (so that the solutions are
assured to be smooth and the preliminary exploration of the $\veceta$
space is faster). This is followed by a second loop in which the best
$\veceta$-model found so far is fixed, and we optimize only for
$\lambda$. The alternate loop procedure is iterated until the maximum
for the evidence is reached. Our tests show that the hyperparameters
generally remain very close to the values found at the second
loop. Hence, this approximation (which is somewhat equivalent to a
line-search minimization along -- alternatively -- the
$\veceta$-parameters and the hyperparameters) works satisfactorily,
significantly reducing the number of iterations necessary to reach
convergence.

\subsubsection{MCMC and Downhill-Simplex Optimization}

The maximization of the evidence is, in general, not an easy task,
because obtaining the function value for a given model can be very
time consuming and the gradient of the evidence over the non linear
parameters $\veceta$ is in general not known or too expensive to
calculate. We have therefore tailored a hybrid optimization routine
which combines a (modified) Markov Chain Monte Carlo (MCMC) search
with a Downhill-Simplex method \citep[see e.g.][for a description of
both techniques]{NR.92}. The preliminary run of the simplex method
(organized in the loops described above) is the crucial step, because
even when launched from a very skewed and unrealistic starting point
$\veceta_{0}$, it usually allows one to reach a good local maximum and
from there to recover the ``best set'' of parameters, i.e. the
$\veceta$-set which corresponds to the absolute maximum of the
evidence function.

The outcome of this first-order optimization is then used as the
starting point for a modified MCMC exploration of the surrounding
evidence surface with an acceptance ratio based on the evaluation of
the normalized quantity $(\evid_{k} - \evid_{\rm bmax})/\evid_{\rm
bmax}$ (where $\evid_k$ is the evidence of the considered point and
$\evid_{\rm bmax}$ is the evidence of the best maximum found so far;
higher maxima are always accepted). When a point is ``accepted'', a
fast Simplex routine is launched from that point to determine the
evidence value of the local maximum to which it belongs. If this turns
out to be a better (i.e.\ higher) maximum than the best value found so
far, it becomes the new starting point for the MCMC. In practice, the
whole procedure can be sped up by about 1 order of magnitude if some
phases of the MCMC exploration and the subsequent local peak climbing
are limited to the optimization of the evidence of lensing only (whose
evaluation is much faster than the calculation of the total evidence),
while the evidence of dynamics provides the decisive criterion to
discriminate between the obtained set of local maxima of the lensing
evidence surface.

As for the implementation of the Downhill-Simplex method,
we have made use of both the freely available optimization packages
MINUIT \citep[][]{minuit} and APPSPACK, a parallel derivative-free
optimization package \citep[see][for an exhaustive description of the
algorithm]{GrKo05, Ko05}, receiving similar results.
The effectiveness of this hybrid approach for our class of problems will
be demonstrated in Sect.~\ref{test.nlin}, where a test case is
considered. 

\section{The Method}
\label{outline}

In this Section we describe the general outline of the framework for
joint gravitational lensing and stellar dynamics analysis (see
Fig.~\ref{scheme} for a schematic flow chart of the method). We
develop an implementation of this methodology (the
{\dynlen}\footnote{Combined Algorithm for Unified Lensing and Dynamics
\mbox{ReconstructiON}.} algorithm) which applies specifically to
axisymmetric potentials and two-integral stellar distribution
functions. This restriction is a good compromise between extremely
fast lensing plus dynamical modeling and flexibility.  Fully triaxial
and three-integral methods \citep[e.g.][]{Cret.99}, although possible
within our framework, are not yet justified by the data quality and
would make the algorithms much slower. Of course, the true
potential might be neither axisymmetric nor even triaxial. Hence, if
significative departures from the assumption of axisymmetry are
present in the lens galaxy, it will have an effect on the lensing and
stellar kinematic reconstructions, which cannot be accounted for by
these simple models. In that case, the model needs to be revised
correspondingly. However, if the model fits the data, in a Bayesian
sense, such that no signicant residuals are left, then a more complex
model is not yet warranted by the data. We emphasize that this does
not prove that the galaxy is not triaxial, has three integrals of
motion, or is even more complex, but only that the data cannot make a
statement about it \citep[see e.g.][for an illustrative discussion on
this topic]{Liddle.07}.

The technical details and a more exhaustive explanation of the
terminology are given in the following Sections and in the Appendices.

First, consider a set of observational data for an elliptical lens
galaxy consisting of (1) the surface brightness distribution of the
lensed images (which we will refer to as the \emph{lensing data}
$\vecd$) and (2) the surface brightness distribution and the first
and second velocity moments map of the lens galaxy itself (hereafter
the \emph{kinematic data} $\vecp$). It is also assumed that the
redshift of the source ($\zs$) and of the lens ($\zd$) are known.

Second, we choose a family of gravitational potentials $\Phi(\vecx,
\veceta)$ which we expect to provide a reasonably good description of
the true potential of elliptical galaxies for some set of model
parameters $\veceta$, such as the normalization constant, core radius,
oblateness, slope, etc.\footnote{We note that in principle $\veceta$
could even be the potential values on a grid of $(R,z)$ and, hence, be
grid-based itself.} The vector $\veceta$ can also include other ``non
intrinsic'' quantities, such as the inclination angle, the position
angle and the coordinates of the center of the elliptical galaxy.

In order to understand the physical characteristics of the lens
galaxy, the first problem consists of finding the specific values of
parameters $\veceta$ that optimize some penalty function based on the
mismatch between the data and the model, within the solution space
allowed by the chosen potential model. A more general and difficult,
but also much more interesting problem, is an objective and
quantitative ranking of a set of different families of potentials
$\{\hat{\Phi}(\vecx, \hat{\veceta})\}$. Our method is designed to
address both of these questions; given the data and a choice for the
potential, the algorithm yields the most likely solution of linear
parameters and a value for Bayesian evidence $\evid(\veceta)$. The
maximum evidence solution $\evid_{\rm max}(\veceta_{\rm max})$ allows
direct and objective model family comparison. The model family can
also include different choices of pixel scales, regularization, PSF
models, etc.

A comparison based on the evidence, as was described in
Section~\ref{bayes}, overcomes many of the traditional hurdles in
regularization and objective model comparison \citep[see also
e.g.][for a thorough discussion of the Bayesian framework in the
context of non parametric lensing]{Suyu.06, Mars.06} and is, therefore,
extremely powerful.

\subsection{The {\dynlen} algorithm}
\label{algorithm}

Whereas the lensed-image reconstruction is general and can
straightforwardly be applied to an arbitrary potential, in the case of
the dynamical modeling we describe how it can be coherently integrated
into the general framework, but then focus on the implementation for
axisymmetric potentials and two-integral distribution functions.
The algorithm requires a set of observables $\vecd$ and $\vecp$, and
the choice of a parametric model $\Phi(\vecx, \veceta)$. An initial
set $\veceta_{i}$ (with $i = 0$) for the non linear parameters is
selected. We indicate the corresponding potential as $\Phi_{i}
\equiv \Phi(\vecx,\veceta_{i})$. This potential is used for both the
lensed-image reconstruction and the dynamical modeling, so that the
method is completely self-consistent and makes full use of all the
data constraints.

\subsubsection{Gravitational Lens Modeling}

The basic idea for the lensed-image reconstruction is the following.
(1) One describes the source brightness distribution by a set of
(possibly irregularly spaced) pixels in the source plane $\vec s$
with each pixel value representing the source surface brightness at
the location of the pixel. (2) A matrix operator ~$\matL$ is then
constructed for a given lens potential, which multiplied by a given
$\vec s$ (and blurred by the PSF) represents the observed lensed image
\citep[see e.g.][]{Warr.03, Treu.04, Koop.05, Dye.05, Suyu.06, Wayth.06,
Brewer.06a}.

In practice, the three-dimensional potential $\Phi_{i}$ is integrated
along the line of sight $z'$ to obtain the projected potential
$\psi_{i}$. The deflection angle is then determined from
$\psi_{i}$. The particular grid-based source reconstruction method
introduced in \citet{Treu.04} and \citet{Koop.05} then allows one to
quickly construct the lensing operator $\matL_{i} \equiv
\matL(\veceta_{i})$ and to determine the most probable pixelized
source $\vecs_{\rm MP, i}$ by maximizing the posterior probability
(Section~\ref{lin.opt}).  More discussion of the lensing routine is
presented in Section~\ref{len}.

\subsubsection{Stellar Dynamical Modeling}

To construct the axisymmetric dynamical model (and the corresponding
operator $\matQ$, see Sect.~\ref{linear.optim}) that reproduces the
kinematic data set, a Schwarzschild method \citep{Schw.79} is used.
Within this flexible and powerful framework, a library of stellar
orbits is integrated in an arbitrary potential $\Phi(\vecx, \veceta)$
and the specific weighted superposition of orbits is determined which
best reproduces the observed surface brightness distribution and
line-of-sight velocity moments of the galaxy \citep[e.g.][]{Rich.80,
Rich.84}.

In the case of the {\dynlen} algorithm for axisymmetric potentials
$\Phi(R,z,\veceta)$, we use the two-integral Schwarzschild method
developed by \citet{Cret.99} and \citet{Vero.02}. In contrast to the
``classical'' Schwarzschild method, here the building blocks for the
galaxy are constituted not by orbits, but by ``two-integral
components'' (TICs), derived from Dirac $\delta$ two-integral
distribution functions (i.e.\ as functions of the orbital energy $E$
and the axial component of the angular momentum $L_{z}$). A TIC can be
thought of as an elementary building block of toroidal shape,
characterized by a combination of orbits that produces a $1/R$ radial
density distribution and very simple (analytically calculable)
velocity moments. For any TIC the projected observables (e.g.\ in our
case surface brightness and line-of-sight first and second velocity
moments) can then be straightforwardly calculated
given~$\Phi_{i}$. The projected axisymmetric density distribution and
velocity moments can be obtained as a weighted superposition of TICs
\citep{Vero.02}.

The aim of the dynamical modeling is therefore to recover a set of
weights which describe how the stellar orbits (or, in the case of the
axisymmetric implementation, the TIC observables) are linearly
superposed to match the data. In analogy to the lensing case, this is
done by maximizing the posterior probability (Sect.~\ref{lin.opt}).  For
a more extended description of the dynamics routine and the generation
of the TICs, we refer to Section~\ref{dyn} and Appendix~\ref{MCMC}.

\subsubsection{Linear optimization of the posterior probability}
\label{linear.optim}

A consequence of the previous considerations and an important feature
of the algorithm is that both the gravitational lensing source
reconstruction and the dynamical modeling can be expressed in a
formally analogous way as sets of coupled linear equations of the form

\begin{equation}
\label{Ax=b}
\left\{
\begin{array}{ll}
\matL \vecs = \vecd \qquad & \textrm{(lensing)} \\
& \\
\matQ \vecgam = \vecp \qquad & \textrm{(dynamics).} \\
\end{array}
\right.
\end{equation}

The ``lensing operator'' $\matL$ encodes how the source surface
brightness distribution $\vecs$ is mapped on to the observed image
$\vecd$. Each of the $\Ns$ columns of $\matL$ describe how a point
source localized on the corresponding pixel is lensed and possibly
multiple imaged on the image plane grid. Similarly, the ``dynamical
operator'' $\matQ$ contains along its $\Ng$ columns all the
information about the observables generated by each individual orbit
or TIC (i.e.\ the surface brightness and the weighted line-of-sight
velocity moments, written in pixelized form on the same grid as the
data), which are superposed with weights $\vecgam$ to generate the
data set $\vecp$.

The crucial advantage of this formulation lies in the fact that each
element of Eq.~(\ref{Ax=b}) is a linear system of equations which can
be solved in a fast and non iterative way, using the standard linear
optimization techniques. Because both $\matL$ and $\matQ$ are built
from the same lens potential $\Phi(\vec \eta)$, both sets of
equations are coupled through the non linear parameters $\vec \eta$.

Because of the ill-posed nature of the set of equations~(\ref{Ax=b})
(i.e.\ the data are noisy), as was discussed in Section~\ref{lin.opt},
finding the solution for $\vec s$ and $\vec p$ that maximimize the
posterior probability translates into solving a set of (regularized)
linear equations
\begin{equation}
\label{ATA}
\left\{
\begin{array}{l}
(\matLt \CvL \matL + \lambda_{\rm L} \matHt_{\rm L} {\matH}^{}_{\rm L}) 
  \vecs = \matLt \CvL \vecd \\
 \\
(\matQt \CvD \matQ + \lambda_{\rm D} \matHt_{\rm D} {\matH}^{}_{\rm D}) 
  \vecgam = \matQt \CvD \vecp 
\end{array}
\right.
\end{equation}
where $\matC^{}_{\rm L}$ and $\matC^{}_{\rm D}$ are the covariance
matrices for lensing and dynamics, respectively, $\matH^{}_{\rm L}$ and
$\matH^{}_{\rm D}$ are a choice for the regularization matrix, and
$\lambda_{\rm L}$ and $\lambda_{\rm D}$ are the corresponding
regularization hyperparameters. Note that, for $\lambda = 0$ (i.e. in
absence of regularization), the solution of Eqs.~(\ref{ATA}) is
equivalent to the maximum likelihood solution in the case of Gaussian
errors.

Once $\matL$ and $\matQ$ have been constructed, from Eqs.~(\ref{ATA})
we can derive the solutions $\vecs_{\rm MP, i}$ and $\vecgam_{\rm MP,
i}$, relative to the choice of the non linear parameters $\veceta_{i}$
and of the hyperparameters~$\veclam_{i} \equiv (\lambda_{\rm L},
\lambda_{\rm D})_{i}$. This, however, represents just a single
possible model, namely, $\Phi_{i} \equiv \Phi(\veceta_{i})$, belonging
to one family of potentials $\Phi(\veceta)$ and, in general, will not
be the ``best'' model given the data $\vecd$ and $\vecp$.

\subsubsection{Non linear optimization of the Bayesian evidence}

In the framework of the Bayesian statistics, the ``plausibility'' for
each considered model can be objectively quantified through the
evidence (see Section~\ref{bayes}), a merit function which includes
(but is not limited to) a likelihood penalty function, but can also
take into account the effects of the choice of regularization, grid,
PSF, etc. The set of non linear parameters~$\veceta_{\rm best}$, and
the corresponding best model $\Phi_{\rm best} \equiv \Phi(\veceta_{\rm
best})$, is obtained by maximizing the evidence through an iterative
optimization loop. For each cycle $i$, a different set of non linear
parameters $\veceta_{i}$ is chosen, and the most probable solutions
$\vecs_{\rm{MP}, i}$ and $\vecgam_{\rm{MP}, i}$ are found as described
before. The evidence $\evid_{i}$ which characterizes this model is
then calculated to allow an objective comparison between different
models.  

Once the evidence is maximized, we are left with the best model
$\Phi_{\rm best}$ (characterized by the value $\evid_{\rm max}$ for
the evidence) and the best reconstruction for the source brightness
distribution $\vecs_{\rm MP, best}$ and the TIC weights $\vecgam_{\rm
MP, best}$ (which is one-to-one related to the distribution function).
At this point, the algorithm has provided us with the best set of
non linear parameters $\veceta_{\rm best}$, i.e.\ the unknowns that we
are most interested in. Nevertheless, we might wish to tackle the more
general problem of model family comparison, by considering a different
family of potentials $\hat{\Phi}$ and applying again the full
algorithm to it. The result will be a vector $\hat{\veceta}_{\rm
best}$ and a value for the evidence $\hat{\evid}_{\rm max}$ that we
can directly compare to the value $\evid_{\rm max}$ that we had
previously found, determining in this way whether the potential $\Phi$
or $\hat{\Phi}$ represents a better model given the constraints.

We now proceed to describe the lensed-image reconstruction and the
dynamical modeling routines in much greater detail. The reader less
interested in the technical details and more in the application of the
algorithm, can continue with Sect.~\ref{test} without loss of 
continuity.

\begin{figure*}[!t]
\begin{center}
\resizebox{1.00\hsize}{!}{\includegraphics[angle=-90]{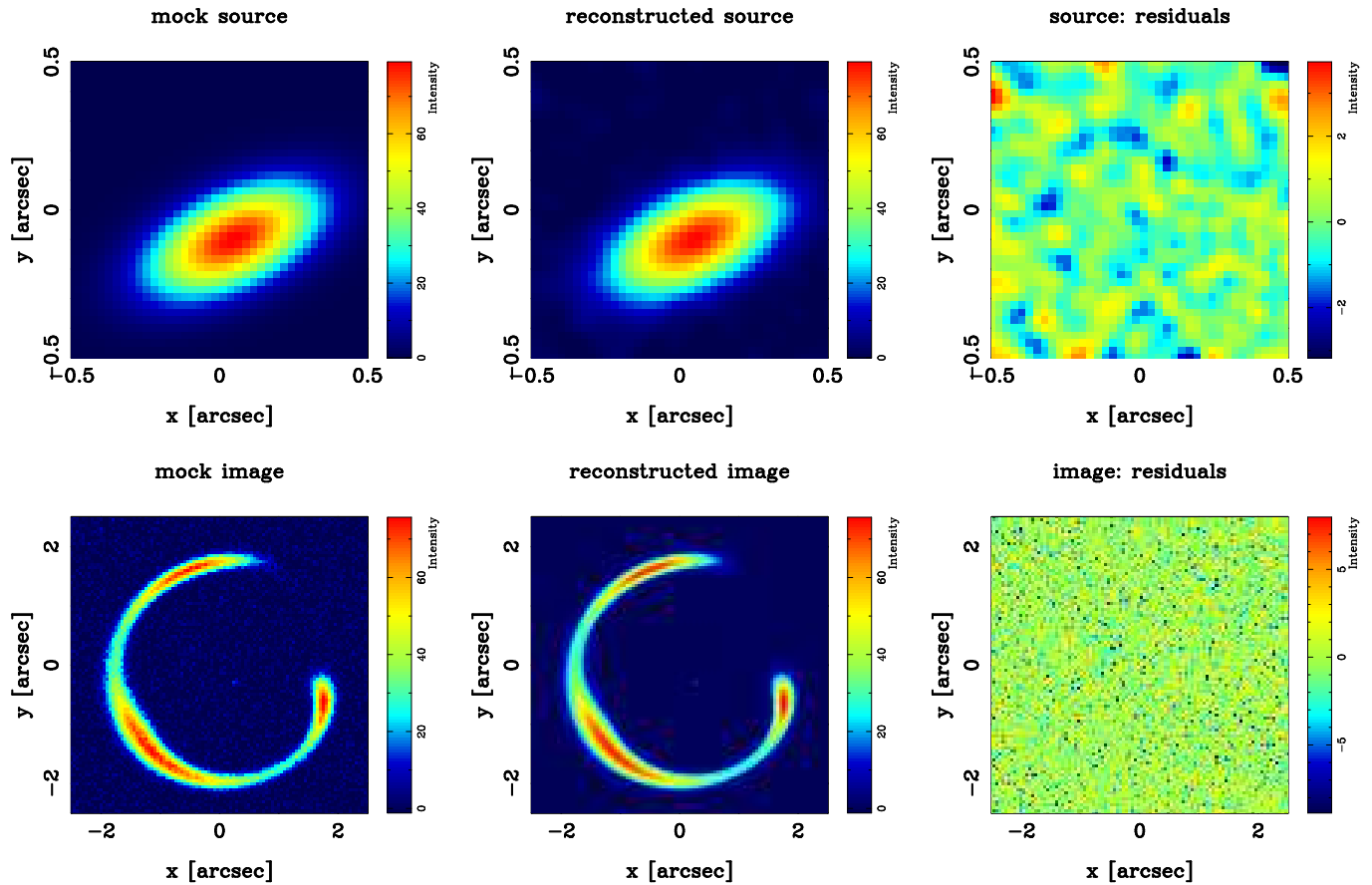}}
\caption{The left columns presents the simulated source $\mocks$
and the corresponding lensed image $\mockd$ with
Gaussian noise added. The middle column shows the reconstructed quantities
obtained with the lensing operator~$\matL$ and a particular choice for
the regularization. The residuals are shown in the right column.}
\label{fig.len.reclin}
\end{center}
\end{figure*}

\section{Gravitational lensing}
\label{len}

Gravitational lensing can be formulated as the reconstruction of an
unknown source brightness distribution $\vecs$ (pixelized on a grid
composed of $\Ns$ pixels) given the observed and PSF-convoluted image
brightness distribution $\vecd$ (sampled on a grid of dimension
$\Nd$). To tackle this problem we have made use of the implementation
of the method of non-parametric source reconstruction initially
developed by \citet{Warr.03} and further refined and/or adapted by
\citet{Treu.04, Koop.05, Dye.05, Suyu.06, Wayth.06, Brewer.06a}

Each pixel $i$ (with $1 \le i \le \Nd$) on the image grid, located at
position $\vecx$, is cast back to the source plane (at position
$\vecy$) through the lensing equation
\begin{equation}
\label{eq.len}
\vecy = \vecx - \vecalp(\vecx)
\end{equation}
where $\vecalp$ is the deflection angle, calculated from the
gravitational potential $\Phi$ as described in
Appendix~\ref{defl.angle}. Since gravitational lensing conserves the
surface brightness $\Sigma$, the equivalence $\Sigma(\vecy) =
\Sigma(\vecx)$ holds (if the effect of the PSF is neglected). In
general, however, $\vecy$ will not exactly correspond to the position of
a pixel of the fixed source grid. Therefore $\Sigma(\vecy)$ is
expressed as a weighted linear superposition of the surface brightness
values at the four pixels $j_{1} \dots j_{4}$ (where the index $j$
runs in the interval $1 \dots \Ns$) which delimit the position
$\vecy$ \citep[see][]{Treu.04}. The weights $w_{1} \dots w_{4}$ for
each of the source pixels are the bilinear interpolation weights
(whose sum adds to unity to conserve flux), and they are stored as the
elements $L_{i j_{1}} \dots L_{i j_{4}}$ of a (very) sparse matrix
$\matL$ of dimension $\Nd \times \Ns$, which represents the
lensing operator. If the image pixel $i$ is cast outside
the borders of the source grid, all the elements of the $i$-th row of
$\matL$ are put to zero. In case we need to be more accurate, we can
split each image grid pixel into $n = n_{1} \times n_{2}$ subpixels
and apply the same procedure as before to construct an $n$-factor
oversampled lensing matrix of dimension $n\Nd \times \Ns$.

The lensing operator is therefore a non linear function of the parameter vector
$\veceta$ through the potential, i.e. $\matL = \matL(\Phi(\veceta))$,
and must be constructed at each iteration of the {\dynlen}
algorithm. From $\matL$ we then construct the \emph{blurred} lensing
matrix $\matM \equiv \matB \matL$, where $\matB$ is a blurring
operator (this is a square matrix of order equal to the number of rows
of~$\matL$) which accounts for the effects of the PSF\footnote{In more
detail, the $i$-th row of the matrix $\matB$ contains a discretized
description of how the PSF is going to spread the surface brightness
at the $i$-th pixel of the image $\vecd$ over the surrounding grid
pixels \citep[see discussion in][]{Treu.04}.}. If we are dealing with
an oversampled lensing matrix, we need to include also a resampling
operator $\matR$ (of dimension $\Nd \times n\Nd$) that sums the
oversampled pixels together so that in the end the blurred lensing
matrix is defined as $\matM \equiv \matR \matB \matL$.

Within this framework, the mapping of the source into the lensed image
can be expressed as the set of linear equations (cfr. Eq.~[\ref{Ax=b}])
\begin{equation}
\label{Ls=d}
\matM \vecs = \vecd .
\end{equation}
As discussed in Section~\ref{lin.opt}, for Gaussian errors the
solution $\vecs_{\rm MP}$ of the ill-conditioned linear system from
equation~(\ref{Ls=d}) is found by minimizing the quadratic penalty
function
\begin{equation}
\label{pen.f.len}
\penf_{\rm len}[\vecs, \Phi(\veceta)] = 
   \frac{1}{2} {(\matM \vecs - \vecd)}^{\rm T} \CvL 
   (\matM \vecs - \vecd) + \frac{\lamlen}{2} 
   {|| \matH \vecs ||}^{2}
\end{equation}
(cf. Eq.~[\ref{pen.f}]) by varying $\vecs$ and finding $d \penf_{\rm
len} / d \vecs = 0$. Here $\matC_{\rm L}$ is the (diagonal) lensing
covariance matrix, $\matH$ is the lensing regularization matrix, and
$\lamlen$ is the corresponding regularization hyperparameter. This
problem translates (see again Sect.~\ref{lin.opt}) into solving the
set of linear equations
\begin{equation}
\label{LTL}
(\matMt \CvL \matM + \lamlen \matHt \matH) \vecs = \matMt \CvL \vecd 
\end{equation}
(cf. Eq.~[\ref{ATA.2}]).

Although in Eqs.~(\ref{pen.f.len}) and~(\ref{LTL}) we have indicated
the regularization matrix simply as $\matH$ for the sake of clarity,
it should be noted that, since $\vecs$ represents a two-dimensional
grid, it is necessary in practice to consider regularization both in
the $x$- and $y$-directions, as described, respectively, by matrices
$\matHx$ and $\matHy$. Therefore, the regularization term in
Eq.~(\ref{LTL}) becomes $\lamlen (\matHxt \matHx + h^{2} \matHyt
\matHy) \vecs$, where $h \equiv \Delta x / \Delta y$ is the ratio
between the horizontal and vertical pixel scales in the case of
``curvature regularization'' \citep[see][for a discussion of different
forms of regularization]{Suyu.06}.  For a specific description of how
the actual regularization matrices are constructed, we refer to
Appendix~\ref{reg}.

\begin{figure*}[!t]
\begin{center}
\resizebox{1.00\hsize}{!}{\includegraphics[angle=-90]{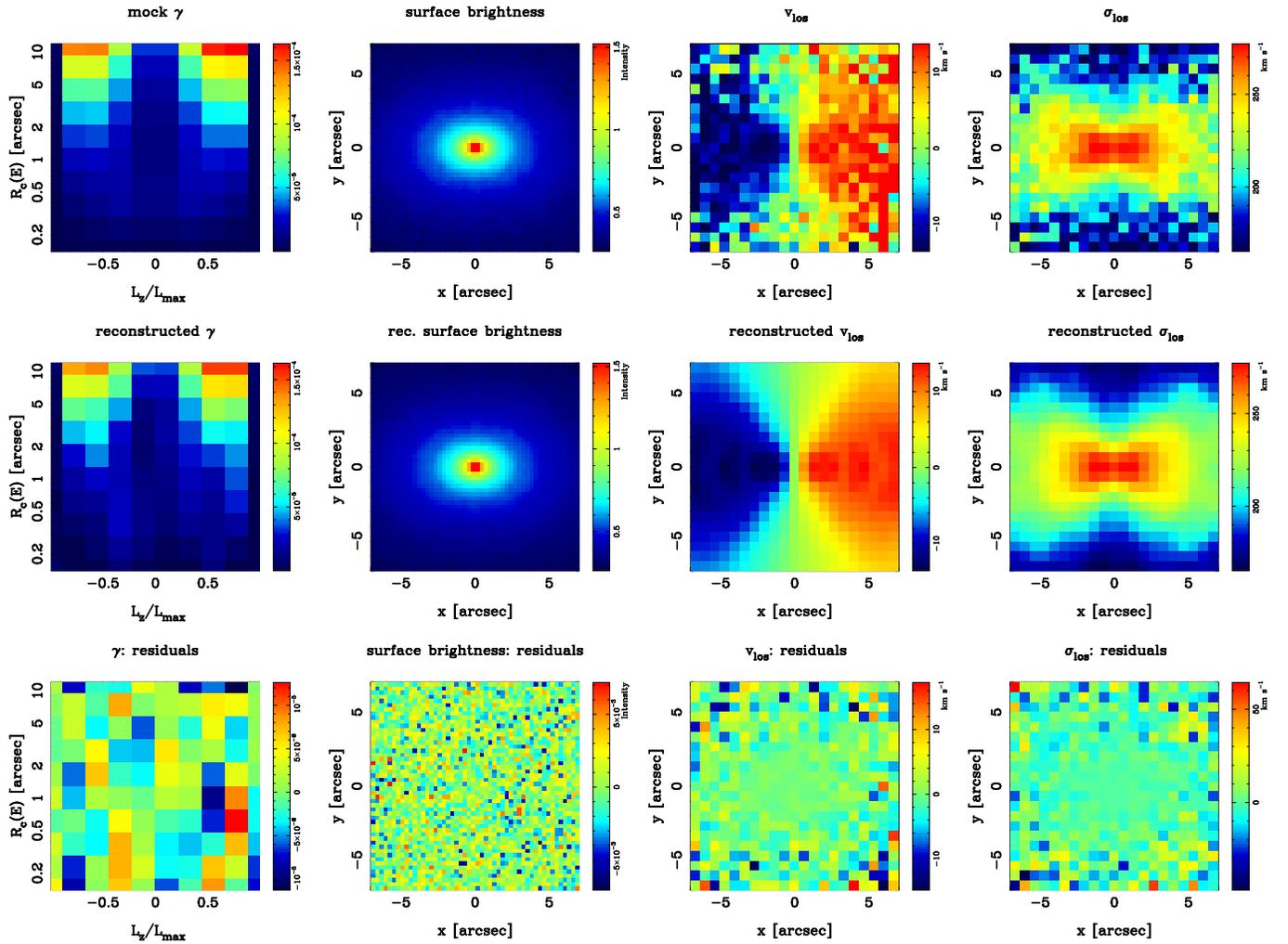}}
\caption{In the first row we display the weighted distribution
function $\vecgam$ (see text) sampled in the integral space $(E,
L_{z})$, together with the set of mock observables generated by this
choice (with non-uniform Gaussian noise added): the surface brightness
distribution $\Sigma$, the line-of-sight stellar streaming velocity
$\vz$ and the line-of-sight velocity dispersion $\sigma^{2}$. The
second row shows the corresponding reconstructed quantities. The
residuals are given in the last row.}
\label{fig.dyn.reclin}
\end{center}
\end{figure*}

\section{Dynamics}
\label{dyn}

In this Section we describe the details of the fast two-integral
Schwarzschild method for the dynamical modeling of axisymmetric
systems, which is implemented in the {\dynlen} algorithm. It should be
noted, however, that Eqs.~(\ref{Qg=p})-(\ref{QTQ}) are also valid in
the general case of arbitrary potentials, provided that $\vecgam$ is
interpreted as the vector of the weights of the different (numerically
integrated) stellar orbits of some orbit library. However, the actual
construction of the ``dynamical operator''~$\matQ$, as described in
Sect.~\ref{Q}, is specific to this particular implementation.

As already shown in Sect.~\ref{algorithm}, from a formal point of view
the problem of dynamical modeling is identical to that of lensing
(Sect.~\ref{len}), essentially consisting of finding the most probable
solution $\vecgam_{\rm MP}$ for the ill-constrained linear system
\begin{equation}
\label{Qg=p}
\matQ \vecgam = \vecp 
\end{equation}
(cf. Eqs.[\ref{Ls=d}] and~[\ref{Ax=b}]). Here $\matQ$ is the
dynamical operator which is applied to the vector $\vecgam$ of the
weights of the building-block $\delta$ distribution functions $\DF$
(the TICs) to generate the set of observables $\vecp$. See
Sect.~\ref{Q} for a more in-depth description of the meaning of the
mentioned quantities and the construction of the matrix $\matQ$. As
described in Section~\ref{lin.opt}, one derives the solution
$\vecgam_{\rm MP}$ by minimizing the quadratic penalty function
\begin{eqnarray}
\label{pen.f.dyn}
\penf_{\rm dyn}[\vecgam, \Phi(\veceta)] 
   & = &
   \frac{1}{2} {(\matQ \vecgam - \vecp)}^{\rm T} \CvD 
   (\matQ \vecgam - \vecp) + \nonumber\\
   & &
   \frac{1}{2} \left( 
   \lamx {|| \matKx \vecgam ||}^{2} +
   \lamy {|| \matKy \vecgam ||}^{2}
   \right) ,
\end{eqnarray}
which corresponds to solving the set of linear equations
\begin{equation}
\label{QTQ}
(\matQt \CvD \matQ + \lamx \matKxt \matKx + \lamy \matKyt \matKy)
\vecgam = \matQt \CvD \vecp 
\end{equation}
(note the equivalence with Eqs.~(\ref{pen.f.len}) and~(\ref{LTL}) for
lensing). Here we have indicated the (diagonal) dynamics covariance
matrix as $\CvD$ and the regularization matrices along the $E$- and
$\Lz$-axes of the $\vecgam$-grid as $\matKx$ and $\matKy$,
respectively. The regularization along these two directions are
(assumed to be) uncorrelated, and the corresponding hyperparameters
$\lamx$ and $\lamy$ must therefore be considered independently.

With respect to their physical meaning, however, the construction of
the linear operators $\matL$ and $\matQ$ is a markedly distinct
problem: the lensing operator describes the distortion to which the
source surface brightness distribution is subjected, while the
dynamics operator is effectively a library which stores the projected
observables (surface brightness distribution and unweighted velocity and
velocity dispersion maps) associated with the set of elementary
(i.e. Dirac $\delta$) two-integral distribution functions.

\subsection{The ``dynamics operator'' by means of a two-integral 
axisymmetric Schwarzschild method}\label{Q}

In this Section we describe how to construct the dynamics operator
$\matQ$ (which is the most complex part of implementing
Eq.~[\ref{QTQ}] explicitly), introducing a new and extremely fast
Monte Carlo implementation of the two-integral components
Schwarzschild method, as proposed by \citet{Cret.99} and
\citet{Vero.02}. The Schwarzschild method is a powerful and flexible
numerical method to construct numerical galaxy models without having
to make any prior assumptions regarding the shape, the symmetry, the
anisotropy of the system, etc.

In its original implementation \citep{Schw.79}, the procedure works as
follows. An arbitrary density distribution (possibly motivated by
observations) is chosen for the galaxy and the corresponding potential
is computed by means of the Poisson equation. One then calculates a
library of stellar orbits within this potential and finds the specific
superposition of orbits which reproduces the initial density
distribution. The method can be generalized to also treat cases in
which density and potential are not a self-consistent pair and to
include kinematic constraints \citep[see e.g.][]{Rich.80, Rich.84,
Pfen.84, Rix.97}.

Orbits, however, are not the {\sl only} possibility for the building
blocks of a Schwarzschild method if one only aims to construct
two-integral axisymmetric models for galaxies. Given an axisymmetric
potential $\Phi(R,z)$, one can also consider more abstract
constituents called two-integral components or TICs
\citep[see][]{Cret.99, Vero.02} which correspond to elementary
Dirac $\delta$ distribution functions completely specified by a choice
of energy $E_{j}$ and angular momentum $L_{z,j}$,
\begin{equation}
\label{TIC}
f_{j}(E_{j}, L_{z,j}) = \left\{
\begin{array}{ll}
\displaystyle \frac{\Cj}{2} \delta(E - E_{j}) \delta(L_{z} - L_{z,j}) 
& \textrm{inside ZVC}\\
& \\
0 & \textrm{elsewhere}
\end{array}
\right.
\end{equation}
where $\Cj \equiv \mathcal{C}_{[E_{j}, L_{z,j}]}$ is a normalization
coefficient chosen such that all the TICs have equal mass (see
Appendix~\ref{normalization} for an explicit expression for
$\Cj$). The zero-velocity curve (ZVC) is the curve in the meridional
plane $(R,z)$ for which
\begin{equation}
\label{E=T+U}
E_{\rm kin} = V(R,z) - \frac{L_{z}^{2}}{2 R^{2}} - E = 0 ,
\end{equation}
where $E$ is the relative energy and $V(R,z) \equiv -\Phi(R,z)$ is the
relative potential; another frequently useful quantity is the
effective potential $V_{\rm eff}(R,z) = V(R,z) - L_{z}^{2} / 2 R^{2}$.

A TIC-based Schwarzschild method has two main advantages. First, the
$j$-th TIC can be interpreted as a particular combination of all
orbits (both regular and irregular, see
\citeauthor{Cret.99}~\citeyear{Cret.99}) with energy $E_{j}$ and
angular momentum $L_{z,j}$ which can be integrated in the potential
$\Phi(R,z)$ and completely fill the ZVC. Therefore, the TICs
constitute a family of building blocks smoother than the regular
orbits (which may have sharp edges) and automatically take into
account the irregular orbits. Second, the unprojected density and
velocity moments for the TICs have simple analytic expressions, which
makes this method much faster than the ordinary orbit integration.

From the definition from equation~(\ref{TIC}), the density
distribution in the meridional plane generated by the $j$-th TIC is
given by \citep[see][]{Binn.book}
\begin{equation}
\label{TIC.rho}
\rho_{j}(R,z) = \left\{
\begin{array}{ll}
\displaystyle
\frac{\pi \Cj}{R} & \textrm{inside ZVC}\\
& \\
0 & \textrm{elsewhere,}
\end{array}
\right.
\end{equation}
while the only non-zero velocity moments have the following
expressions inside the ZVC:
\begin{equation}
\label{TIC.vphi}
\rho_{j} \vphi_{j} = \frac{\pi \Cj}{R^{2}} L_{z,j} ,
\end{equation}
\begin{equation}
\label{TIC.vqphi}
\rho_{j} \vqphi_{j} = \frac{\pi \Cj}{R^{3}} L_{z,j}^{2} ,
\end{equation}
\begin{equation}
\label{TIC.vz}
\rho_{j} \vqR_{j} = \rho_{j} \vqzz_{j} =  
\frac{\pi \Cj}{R} \left[ V_{\rm eff} (R,z) - E_{j} \right] ,
\end{equation}
and they vanish elsewhere.

Note that we cannot directly compare the quantities described by
Eqs.~(\ref{TIC.rho})-(\ref{TIC.vz}) with the observations. Before this
can be done, we need to calculate the projected quantities, grid them,
and convolve them with the appropriate PSF (and possibly regrid them
again in the case of subgridding; see text).

The surface brightness $\SB_{j}$ (sampled on a grid of $\nSB$
elements) and the weighted line-of-sight velocity moments $\SB_{j}
\vz_{j}$ and $\SB_{j} \vqz_{j}$ (both sampled on a grid of $\nvz$
elements) can be obtained semi-analytically through multi-dimensional
integrals \citep[refer to][]{Vero.02}. Because the same semi-analytic
approach to calculate $\matQ$ is rather time consuming, we have
developed a very efficient Monte Carlo implementation of the
two-integral Schwarzschild method, detailed in Appendix~\ref{MCMC},
which is several orders of magnitude faster.

Whatever technique is adopted, the projected and PSF-convoluted
quantities for the $j$-th TIC, sampled on the appropriate grids,
constitute the $j$-th column of the $\matQ$ matrix. Therefore, if the
galaxy model is built using a library of $\Ng = \nE \times \nLz$ TIC
building blocks (where $\nE$ and $\nLz$ are the number of samplings,
respectively, in energy and angular momentum), the dynamics operator
$\matQ$ turns out to be a dense matrix of dimensions $\Ng \times
(\nSB + 2 \nvz)$. 

The meaning of the vector $\vecgam$ in Eq.~(\ref{Qg=p}) is now clear:
its $\Ng$ dimensionless non negative elements $\gamma_{j}$ describe the
weights of the linear superposition of the model observables generated
by the library of TICs, and we look for the solution $\vecgam_{\rm
MP}$ (given by Eq.~[\ref{QTQ}]) which best reproduces the data set
$\vecp$. Moreover, as explained in Appendix~\ref{normalization}, the
weights $\gamma_{j}$ are proportional to the light contributed by the
TIC to the galaxy and related to the reconstructed dimensional
distribution function $\textrm{DF}(E_{j}, L_{z,j})$ when they are
normalized with the TIC area $\Azvc$ in the meridional plane and the
surface of the cell $\AELz$ in integral space.

\section{Demonstration of the method}
\label{test}

In this Section we describe several of the tests that we have
performed to show the proper functioning of the method.  We construct
a simulated data set (including realistic noise) for both lensing and
dynamics by adopting the potential given in Eq.~(\ref{evans.pot})
\citep{Evans.94a} and a particular choice for the non linear
parameters $\veceta$ (see Sect.~\ref{test.setup} for a description of
the setup). Then in Sect.~\ref{test.lin}, keeping the non linear
parameters fixed, we test the linear optimization part of the method
by showing that the code is able to faithfully reconstruct the source
surface brightness distribution and the distribution function used to
generate the mock data. Finally, in Sect.~\ref{test.nlin} we execute a
full run of the code. We use the mock set of simulated data as input
and adopt a starting set of non linear parameters considerably skewed
from the ``true'' values, in order to verify how reliably the linear
and non linear parameters are recovered. Note that, for conciseness,
what we refer to as the ``evidence'' $(\evid)$ is, rigorously
speaking, the logarithm of the evidence as presented in
Sect.~\ref{bayes}. See Appendix~\ref{evidence} for its precise
definition.

\subsection{The test setup}
\label{test.setup}

As mentioned, for testing purposes it is convenient to make use of the
Evans' power-law potentials \citep{Evans.94a, Evans.94b}
\begin{equation}
\label{evans.pot}
\Phi(R,z) = - \frac{\Phiz \Rcore^{\beta}}
            {\left( \Rcore^2 + R^2 + \displaystyle \frac{z^2}{q^2} 
            \right)^{\beta/2}} \qquad \beta \neq 0 ,
\end{equation}
where $\Phiz$ is the central potential, $\Rcore$ is a core radius and
$q$ is the axis ratio of the equipotentials. For $\beta = 0$ this
becomes the \citet{Binn.81} logarithmic potential.

What makes this class of axisymmetric galaxy models suitable for the
setup of a test case, is the overall simplicity of their
properties. The power-law galaxy models are characterized by
elementary two-integral distribution functions and provide fully
analytical expressions for the density distribution, associated with
the potential via the Poisson equation, the second velocity moments
and the deflection angle (see Appendix~\ref{evans}). Moreover, even
the projected quantities (i.e.\ the projected surface density and the
line-of-sight second velocity moment) are analytical. The mean stellar
streaming velocity $\vphi$ is assigned through a physically motivated
prescription \citep[see Sect.~2.3 of][]{Evans.94b} which leads to a
simple odd part for the distribution function, which does not
contribute to the density and the second velocity moment.

With the choice of the potential from equation~(\ref{evans.pot}) for
the lens galaxy, the elements of the $\veceta$ vector of non linear
parameters are $\beta$, $q$, $\Rcore$ and $\Phiz$ (or equivalently,
through Eq.~[\ref{lens.str}], the lens strength $\talp$). In addition,
$\veceta$ includes the parameters that determine the observed geometry
of the system: the inclination~$i$, the position angle~$\PA$, and the
coordinates~$\vecxi_{0}$ of the center of lens galaxy on the sky
grid. We note that $\veceta$ can also include the external shear
strength and position angle, although in our current tests we assume
negligible shear for the sake of simplicity and to fully concentrate
on the parameters of the lens galaxy only, in order to test whether
their degeneracies can be broken by the combination of lensing and
stellar kinematic data. We also need to make a choice for the size of
the grids in pixels. This actually constitutes an explicit prior
choice (just like the type of regularization). The evidence however
can be used to test exactly these types of assumptions. Explicitly, our
test setup is the following.

\begin{itemize}

\item[$\bullet$] \emph{Lensing:} For the test setup, we adopt a $40
  \times 40$ pixel grid ($\Ns = 1600$) in the source plane and a $100
  \times 100$ pixel grid ($\Nd = 10000$) in the image plane. The
  lensing operator is built using an oversampling factor of~$3$ to
  improve the quality of the reconstruction.

\item[$\bullet$] \emph{Dynamics:} In the two-integral phase space we
  consider a grid of $\nE = 10$ elements logarithmically sampled in
  $\Rc(E)$ and $\nLz = 5$ elements linearly sampled in angular
  momentum, i.e.\ a total of $\Ng = 100$ TICs (note that the grid must
  be mirrored for negative $L_{z}$). Each TIC is populated with $\nTIC
  = 10^{5}$ particles. This number of TICs (when compared to the grid
  size for the lensing source for example) has been verified to be a
  good compromise between the quality of the distribution function
  reconstruction and the heavy computational power needed in the
  iterative scheme for the construction of many TICs. The surface
  brightness and the line-of-sight velocity moments are sampled on
  different grids\footnote{The observed surface brightness and
  line-of-sight velocity moments often are obtained with different
  instruments, hence the need for different grids.}  of, respectively,
  $50 \times 50$ and $21 \times 21$ elements ($\nSB = 2500$ and $\nvz
  = 441$). Analogous to the case of lensing, an oversampling factor
  of~$3$ is adopted in the construction of the operator~$\matQ$.

\end{itemize}

\subsection{Demonstration of the linear reconstruction}
\label{test.lin}

We select and fix the following arbitrary (albeit physically
plausible) set of values for the~$\veceta$ vector: $\beta = 0.28$, $q
= 0.85$, $\Rc = 0.3''$, $\talp = 4.05''$ (and the ratio\footnote{We
indicate as $\Ds$, $\Dd$ and $\Dds$ the angular diameter distance of,
respectively, observer-source, observer-lens and lens-source.}
$\Dds/\Ds$ is taken to be $0.75$), $i = 60^{\circ}$, $\PA = 0^{\circ}$
and $\vecxi_{0} = (0.25'', -0.25'')$. This makes it possible to
construct the lensing operator\footnote{To obtain $\matM$, which is
the \emph{blurred} lensing operator, it is necessary to have the
blurring operator $\matB$, which we construct from the $7 \times 7$
pixel PSF model.}~$\matM$ and the dynamics operator~$\matQ$
(Sects.~\ref{len} and~\ref{dyn}), which are then used to generate a
simulated data set. With a {\machine} machine, the construction of the
sparse lensing matrix is a very fast process (less than 1 s).
Constructing the dynamics operator is more time consuming, although
requiring in the above case still only about 7~s, i.e.\ of the
order of 100~ms per TIC (it should be noted that this is a very
short time for building a dynamical library and is a direct
consequence of the Monte Carlo implementation described in
Appendix~\ref{MCMC}.) In addition:

\begin{itemize}

\item[$\bullet$] \emph{Lensing:} We adopt an elliptical Gaussian
  source surface brightness distribution $\mocks$ and using the Evans'
  potential from equation~(\ref{evans.pot}) (see
  Eq.~[\ref{evans.alpha}] for the analytic expression of the
  corresponding deflection angle) we generate the blurred lensed
  image. From this we obtain the final mock image $\mockd$ by adding a
  Gaussian noise distribution with $\sigma = 0.03$~times the image
  peak value. This is illustrated in the first column of
  Fig.~\ref{fig.len.reclin}.

  The reconstruction of the non-negative source $\recs$, from the
  simulated data $\mockd$, is obtained by solving the linear system of
  equations~(\ref{LTL}), with the adoption of a fiducial value for the
  regularization hyperparameter ($\log \lamlen = -1.0$). Although the
  matrix $\matMt \CvL \matM + \lamlen \matHt \matH$ in Eq.~(\ref{LTL})
  is large ($10000 \times 1600$), it is very sparse and therefore,
  using L-BFGS-B, it only takes $\lesssim 1$ s to find the
  solution. The result is shown in Fig.~\ref{fig.len.reclin}
  (\emph{middle column}) together with the residuals (\emph{right
  column}).

\item[$\bullet$] \emph{Dynamics:} We generate the simulated data set
  $\mockp$ for a self-consistent dynamical system described by the
  distribution function of the Evans' power-law potential. We adopt
  the same parameters used for the above potential. This kind of
  assumption, in general, is not required by the method. However, we
  adopt it here because it immediately and unambiguously allows us to
  check the correctness of the simulated data in comparison to the
  analytic expressions. In this way, it is possible to verify that the
  considered TIC library $\mockgam$, although composed of only $100$
  elements, indeed represents a fair approximation of the true
  distribution function. 

  The first two panels in the first row of Fig.~\ref{fig.dyn.reclin}
  show the TIC weights $\mockgam$ (i.e.\ the weighted distribution
  function) sampled over the two-integral space grid. The remaining
  panels display the simulated data: the surface brightness
  distribution, the line-of-sight streaming velocity and the
  line-of-sight velocity dispersion\footnote{As explained in
  Appendix~\ref{MCMC}, the calculated quantity is not the
  line-of-sight velocity dispersion $\slos$, but instead $\vqz$, which
  is additive and can therefore be directly summed up when the TICs
  are generated and superposed. The line-of-sight velocity dispersion
  to compare with the observation, however, can be simply obtained in
  the end as:
  \begin{equation}
    \label{slos}
    \slos = \vqz - \vz^{2}.
  \end{equation}} 
  As can be seen, non-uniform Gaussian noise has been added to the
  simulated data (its full characterization is taken into account in
  the covariance matrix $\CvD$). The noise on the surface brightness
  is a few percent of the value of this quantity in the inner region,
  while, in order to make the test case more realistic, the noise on
  the velocity moments is increasing outwards and becomes quite severe
  in the external regions.\footnote{The noise on the velocity moments
  (as in the case of real data) depends on the signal-to-noise ratio
  of the surface brightness at a given point. Hence the
  signal-to-noise ratio on the velocity moments decreases outward,
  since the noise is constant but the signal is not. The noise level
  was set at a level generally consistent with both the \emph{HST}
  imaging, Keck spectroscopy and VLT VIMOS-IFU data that is currently
  being obtained as part of the SLACS survey.}\\

  The reconstruction of the non negative TIC weights $\recgam$ is
  given by the solution of Eq.~(\ref{QTQ}) (the chosen values for the
  hyperparameters are $\log \lamx = 9.2$ and $\log \lamy = 9.4$). The
  reconstruction of the dynamical model constitutes, together with the
  generation of the TIC library, the most time consuming part of the
  algorithm, requiring almost $10$ s. This is a consequence of
  the fact that the $3382 \times 100$ dynamics operator $\matQ$,
  although much smaller than $\matL$, is a fully dense matrix. If the
  number of TICs is significantly increased, the time required for
  calculating the term $\matQt \CvD \matQ$ in Eq.~(\ref{QTQ}) and to
  solve that set of linear equations with the L-BFGS-B method
  increases very rapidly (as does the time needed to generate the
  TICs, although less steeply), and therefore the dynamical modeling
  is typically the bottleneck for the efficiency of the method.

  The results of the reconstruction are shown in the second row of
  Fig.~\ref{fig.dyn.reclin} (whereas the third row shows the
  residuals). The reconstruction is generally very accurate.

\end{itemize}

Having verified the soundness of the linear reconstruction algorithms,
the next step is to test how reliably the method is able to recover
the ``true'' values of the non linear parameters $\veceta$ from the
simulated data, through the maximization of the evidence penalty
function $\evid(\veceta)$.

\subsection{Non linear optimization}
\label{test.nlin}

We first run the linear reconstruction for the reference model
$\model_{\rm ref}$ described in Sect.~\ref{test.lin}, optimized for
the hyper-parameters, to determine the value of the total evidence
$\evid_{\rm tot, ref} = \evid_{\rm len, ref} + \evid_{\rm dyn, ref}$
(reported in the first column of Table~\ref{table.evid}). Since this
is by definition the ``true'' model, it is expected (provided that the
priors, i.e.\ grids and form of regularization, are not changed) that
every other model will have a smaller value for the evidence.

\begin{table}
\begin{center}
\caption{Results for iterative search of the best model parameters via
evidence maximization (see text).}
\smallskip
\begin{tabular}{ c c c c c c }
\hline
\noalign{\smallskip}
  & & $\model_{\rm ref}$ & $\model_{0}$ & $\model_{\rm step}$ & $\model_{\rm final}$ \\
\noalign{\smallskip}
\hline
\noalign{\smallskip}
           & $i\;\textrm{[deg]}$ & 60.0   & 25.0   & 75.6  & 62.6   \\
non linear & $\talp$             & 4.05   & 5.60   & 3.86  & 3.99   \\
parameters & $\beta$             & 0.280  & 0.390  & 0.288 & 0.285  \\
           & $q$                 & 0.850  & 0.660  & 0.880 & 0.857  \\
\noalign{\smallskip}
\hline
\noalign{\smallskip} 
           & $\log \lamlen$  & -1.04    &  0.00 & -1.04 & -1.04  \\ 
hyper-     & $\log \lamx$    &  8.00    & 12.0  &  7.99 &  8.00  \\
parameters & $\log \lamy$    &  9.39    & 12.0  &  9.55 &  9.39  \\
\noalign{\smallskip}
\hline
\noalign{\smallskip}
         & $\evid_{\rm len}$ & -22663 & -58151  & -22668 & -22661  \\
evidence & $\evid_{\rm dyn}$ &  12738 & -156948 &  12672 &  12856  \\
         & $\evid_{\rm tot}$ &  -9925 & -215099 &  -9996 &  -9805  \\
\noalign{\smallskip}
\hline
\end{tabular}
\label{table.evid}
\end{center}
{\footnotesize The reference model in the second column is the
``true'' model which generated the simulated data sets, and is shown
for comparison. The first group of rows list the non linear parameters
$\veceta$ which are varied in the loop. The second group shows the
hyperparameters. The last group shows the evidence
relative to the considered model; the contributions of the lensing and
dynamics part to the total evidence are also indicated.}
\end{table}

\begin{figure}
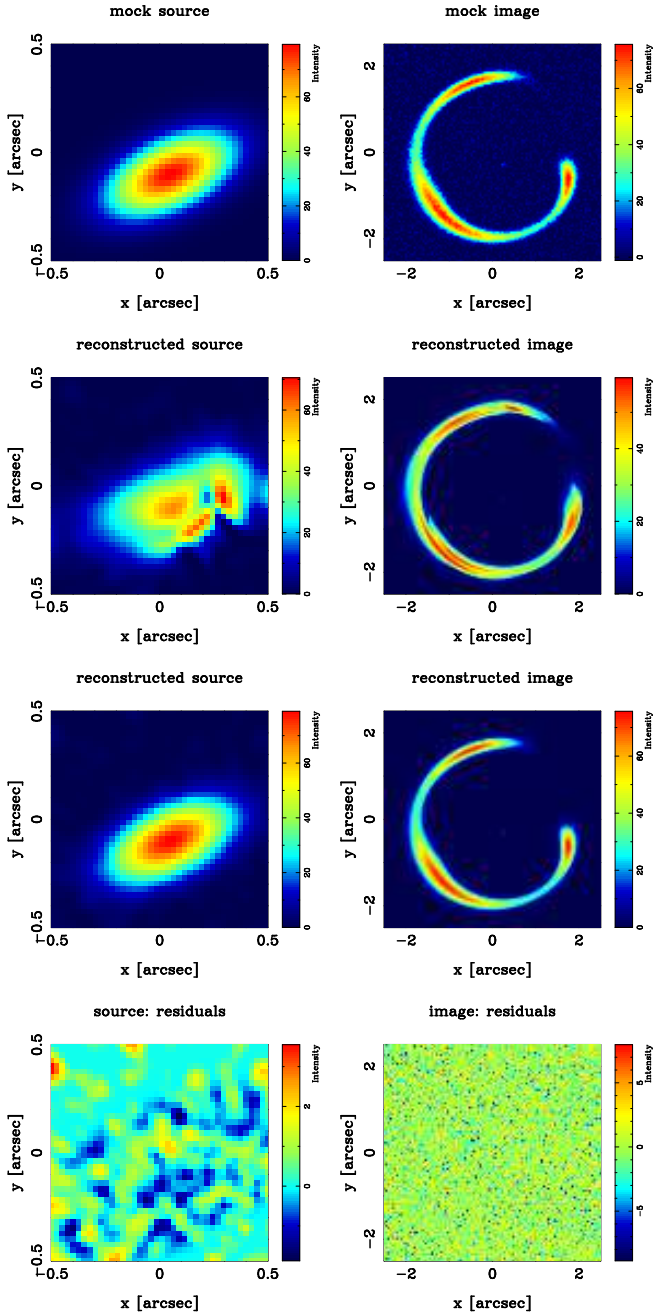

\resizebox{\hsize}{!}{\includegraphics[angle=-90]{LEN.data.ps}}
\vspace{0.0cm}

\resizebox{\hsize}{!}{\includegraphics[angle=-90]{LEN.start.ps}}
\vspace{0.0cm}

\resizebox{\hsize}{!}{\includegraphics[angle=-90]{LEN.final.ps}}
\vspace{0.0cm}

\resizebox{\hsize}{!}{\includegraphics[angle=-90]{LEN.resid.ps}}

\caption{Outcome of the non linear optimization routine for the
lensing block. The first row shows the simulated source and lensed
image. In the second row is the starting model~$\model_{0}$ obtained
with the skewed parameters, which clearly produces a poor
reconstruction of the image (hence the very low value of the
evidence). The third row displays the final model~$\model_{\rm
final}$, that is the ``best model'' recovered through the non linear
optimization for evidence maximization. This should be compared to the
reference model~$\model_{\rm ref}$ (i.e. generated with same $\veceta$
set of parameters which were used to generate the mock data),
presented in the second column of Fig.~\ref{fig.len.reclin}. The last
row shows the residuals of the final model when compared to the
simulated data. Refer to Table~\ref{table.evid} for the non linear
parameters and hyperparameters of the models and the corresponding
value of the evidence.}
\label{fig.len.opt}
\end{figure}

\begin{figure*}[!tb]
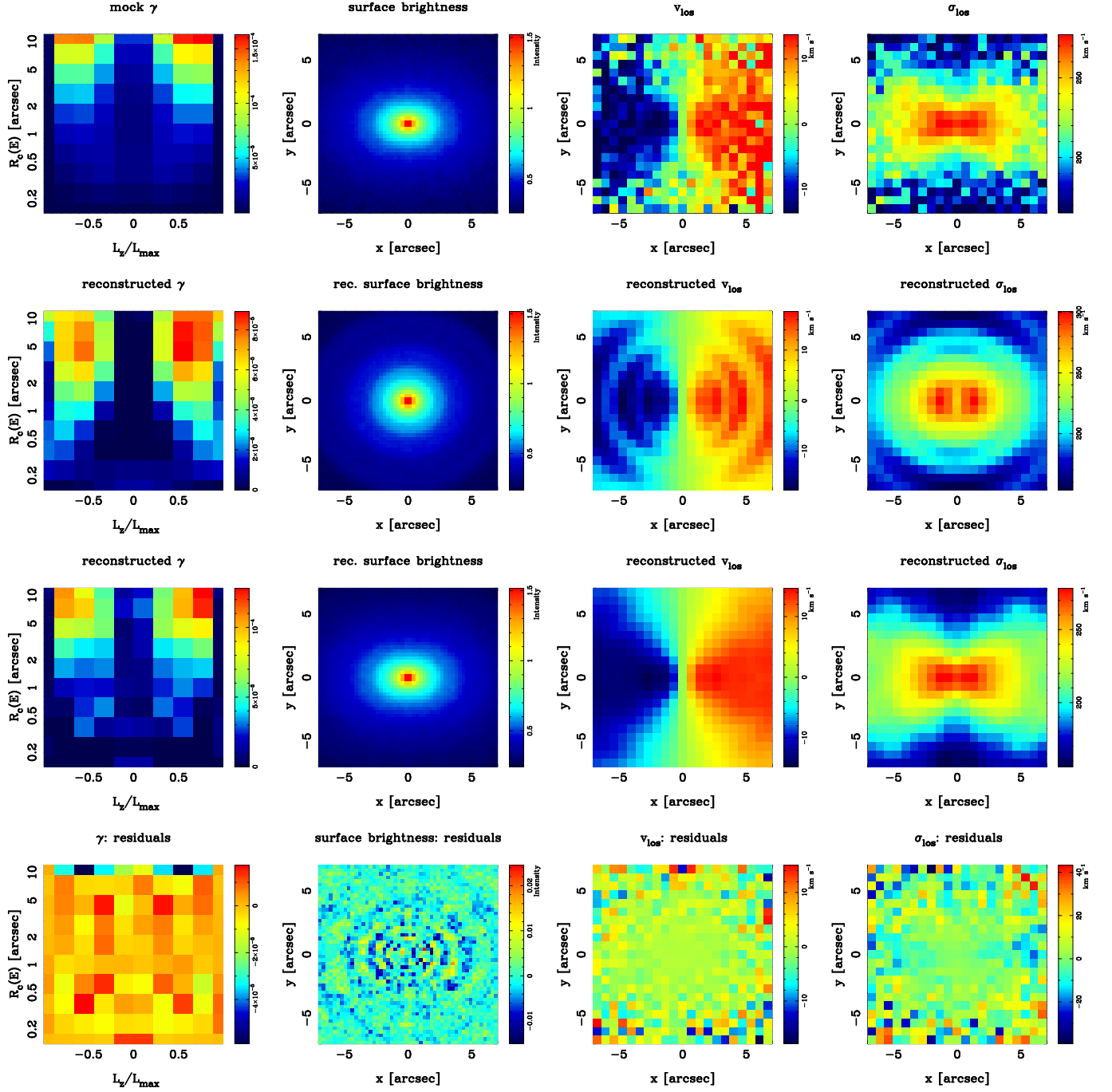
 
\resizebox{\hsize}{!}{\includegraphics[angle=-90]{DYN.data.ps}}
\vspace{0.0cm}

\resizebox{\hsize}{!}{\includegraphics[angle=-90]{DYN.start.ps}}
\vspace{0.0cm}

\resizebox{\hsize}{!}{\includegraphics[angle=-90]{DYN.final.ps}}
\vspace{0.0cm}

\resizebox{\hsize}{!}{\includegraphics[angle=-90]{DYN.resid.ps}}

\caption{Outcome of the non linear optimization routine for the
dynamics block. The first row shows the simulated set of TIC
weights~$\vecgam$ (i.e. the weighted distribution function) in the
integral space and the resulting observables: the surface brightness
$\Sigma$, the line-of-sight streaming velocity $\vz$ and the
line-of-sight velocity dispersion $\slos$.  The second row presents
the reference model~$\model_{\rm ref}$ (i.e. generated with same
$\veceta$ set of parameters which were used to generate the mock
data).  In the second row is the starting model~$\model_{0}$ obtained
with the skewed parameters, which produces a poor reconstruction of
the observables, in particular the velocity moments.  The third row
displays the final model~$\model_{\rm final}$, that is the ``best
model'' recovered through the non linear optimization for evidence
maximization. This should be compared to the reference
model~$\model_{\rm ref}$ (i.e. generated with same $\veceta$ set of
parameters which were used to generate the mock data), presented in
the second row of Fig.~\ref{fig.dyn.opt}.  The last row shows the
residuals of the final model when compared to the simulated data.
Refer to Table~\ref{table.evid} for the non linear parameters and
hyperparameters of the models and the corresponding value of the
evidence.}
\label{fig.dyn.opt}
\end{figure*}

Second, we construct a ``wrong'' starting model $\model_{0} \equiv
\model(\veceta_{0})$ by significantly skewing the values, indicated in
the second column of Table~\ref{table.evid}, of the four non linear
parameters: the inclination~$i$, the lens strength~$\talp$, the
slope~$\beta$ and the axis ratio~$q$. Figures~\ref{fig.len.opt}
and~\ref{fig.dyn.opt} show that $\model_{0}$ is clearly not a good
model for the data in hand. We do not set boundaries on the values
that the non linear parameters can assume during the iterative search,
except for those which come from physical or geometrical
considerations (i.e.  inclination comprised between edge-on and
face-on, $\talp \ge 0$, $0 \le q \le 1$, $0 < \beta \leq 1$). The
position angle, in general, is reasonably well-constrained
observationally\footnote{As a consequence of the assumption of
axisymmetry, the position angle of the total potential should match by
construction the position angle of the surface brightness
distribution.}, and therefore including it with tight constraints on
the interval of admitted values would only slow down the non linear
optimization process without having a relevant impact on the overall
analysis. It is therefore kept fixed during the optimization. We also
keep the core radius of the mass distribution fixed even though in
general it is not well constrained by the surface brightness
distribution and in real applications it should be varied. It always
remains possible, once the best model has been found, to optimize for
the remaining parameters in case a second-order tuning is required.

Adopting~$\veceta_{0}$ as the starting point for the exploration, the
non linear optimization routine for the evidence maximization is run
as described in Sect.~\ref{nlin.opt}. The model $\model_{\rm step}$
(third column of Table~\ref{table.evid}) is what is recovered after
three major loops (the first and the last one for the optimization of
the varying non linear parameters, the intermediate one for the
hyperparameters) of the preliminary Downhill-Simplex optimization, for
a total of $\sim 1000$ iterations, requiring about~$1.2 \times 10^{4}$
s on a {\machine} machine. From this intermediate stage, the final
model $\model_{\rm final}$ is obtained through the combined
MCMC+Downhill-Simplex optimization routine, in a time of the order of
$12-15$ hours.

Further testing has shown that increasing the number of loops does not
produce relevant changes in the determination of the non linear
parameters nor of the hyperparameters, and therefore extra loops are
in general not necessary. We also note that, in all the considered
cases, the first loop is the crucial one for the determination of
$\veceta$, and the one which requires the largest number of iterations
before it converges. It is generally convenient to set the initial
hyperparameters to high values so that the system is slightly
over-regularized. This has the effect of smoothing the evidence
surface in the $\veceta$-space, and, therefore, facilitates the search
for the maximum. The successive loops will then take care of tuning
down the regularization parameters to more plausible values.

\begin{figure*}[tb!]

\resizebox{\hsize}{!}{\includegraphics[angle=-90]{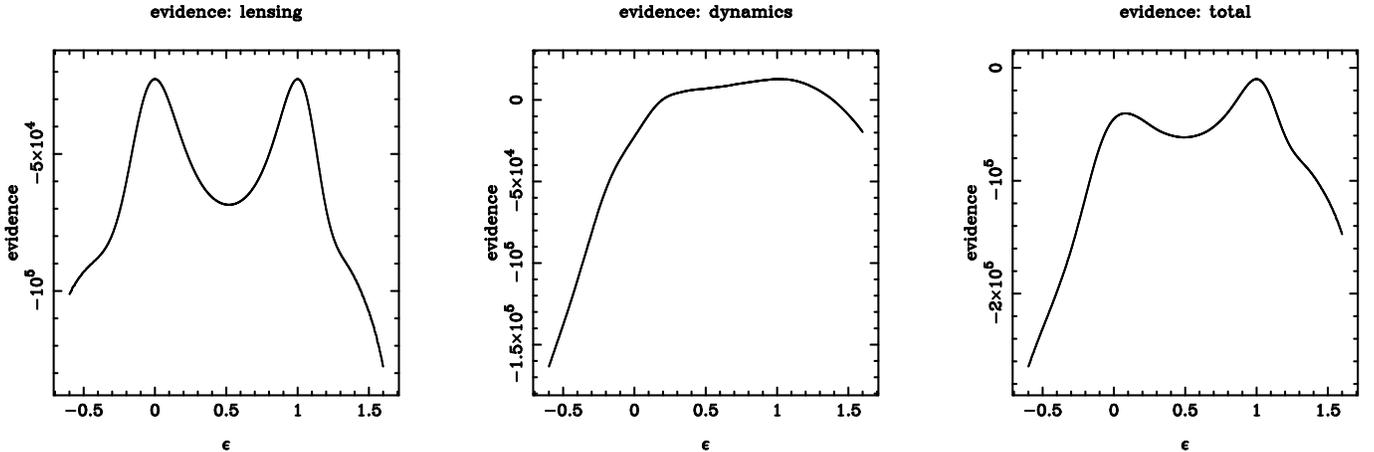}}

\caption{These plots show how the degeneracy between different lens
models is broken when the constraints given by the stellar dynamics
are also considered. The three panels display a cut through the
surfaces of, respectively, lensing evidence, dynamics evidence and
total evidence. The abscissa coordinate $\epsilon$ represents a series
of different $\veceta$ sets, i.e. different models, obtained as a
linear interpolation between the model parameters $\veceta_{\epsilon =
0} \equiv $ ($i = 35\degr$, $\alpha = 5.59$, $\beta = 0.297$, $q =
0.602$), i.e. the best model which is obtained by maximizing for lensing
evidence only, and the model $\veceta_{\epsilon = 1}$ which is the
reference model $\model_{\rm ref}$ of Table~\ref{table.evid}. As far
as only the gravitational lensing is considered, the two maxima in the
evidence are effectively degenerate, with the ``wrong'' model
$\veceta_{\epsilon = 0}$ being slightly preferred ($\Delta \evid_{\rm
len} = 2$). When the contribution of the evidence of dynamics is
considered, however, the degeneracy is broken and the reference model
emerges as indisputably favored by the total evidence ($\Delta
\evid_{\rm tot} \simeq 35000$).}
\label{fig.degen}
\end{figure*}

A comparison of the retrieved non linear parameters for $\model_{\rm
final}$ (last column of Table~\ref{table.evid}) with the corresponding
ones for the reference model reveals that all of them are very
reliably recovered within a few percent (the most skewed one is the
inclination~$i$, being only $\sim 4\%$ different from the ``true'' value). The
panels in the second and fourth rows of Figures~\ref{fig.len.opt}
and~\ref{fig.dyn.opt} clearly show that the two models indeed look
extremely similar, to the point that they hardly exhibit any
difference when visually examined.

The evidence of the final model, when compared with the value for the
reference model, turns out to be higher by $\Delta \evid \equiv
\evid_{\rm final} - \evid_{\rm ref} = 120$, which might look
surprising since $\model_{\rm ref}$ is by construction the ``true''
model. The explanation for this is given by the cumulative effects of
numerical error (which enters mainly in the way the TICs are
generated\footnote{In fact, when in this same case the TICs are
populated with $10$ times the number of particles, i.e. $\nTIC =
10^{6}$, the evidence increases for both $\model_{\rm ref}$ and
$\model_{\rm final}$, and the reference model is now favored (of a
$\Delta \evid \equiv \evid_{\rm ref} - \evid_{\rm final} \simeq 30$)
as expected. Obviously, this comes at a significant cost in
computational time, with the dynamical modeling becoming approximately
an order of magnitude slower.}), finite grid size, and noise (in
particular on the velocity moments), which all contribute to slightly
shift the model. Such a shift is, however, well within the model
uncertainties (Sect.~\ref{error.analysis}) and therefore not
significant.

\subsubsection{Degeneracies and Biases}
\label{biases}

Extensive MCMC exploration reveals that the lensing evidence surface
$\evid_{\rm len}(\veceta)$ is characterized by multiple local maxima
which effectively are degenerate even for very different values of
$\veceta$. Indeed, for the potential Eq.~(\ref{evans.pot}), one can
easily observe from Eq.~(\ref{evans.alpha}) that a relation between
$q$, $i$, and $\Phi_0$ exists that leaves the deflection angles
and, hence, the lens observable invariant. 

In this situation, the dynamics plays a crucial role in breaking the
degeneracies and reliably recovering the best values for the non
linear parameters. A clear example is shown in Figure~\ref{fig.degen},
where cuts along the evidence surfaces are presented. The left panel
of Figure~\ref{fig.degen} displays two almost degenerate maxima in
$\evid_{\rm len}$; indeed, the maximum located at the linear
coordinate $\epsilon = 0$ (corresponding to the set of non linear
parameters $i = 35\degr$, $\alpha = 5.59$, $\beta = 0.297$, $q =
0.602$) has a value for the lensing evidence of $-22661$, and
therefore, judging on lensing alone, this model would be preferred to
the reference model (see Table~\ref{table.evid}) which corresponds to
the other maximum at $\epsilon = 1$. When the evidence of the dynamics
(Fig.~\ref{fig.degen} \emph{central panel}) is considered, however,
the degeneracy is broken and the ``false'' maximum is heavily
penalized with respect to the true one, as shown by the resulting
total evidence surface (Fig.~\ref{fig.degen} \emph{right panel}).

Test cases with a denser sampling of the integral space (e.g. $\nE
\times \nLz = 12 \times 8$, $18 \times 9$, $20 \times 10$) were also
considered. This analysis revealed that, although one obtains more
detailed information about the reconstructed two-integral distribution
function at the cost of a longer computational time (e.g. in the $18
\times 9$ case the loop phase is slower by more than a factor of 2),
the accuracy of the recovered non linear parameters -- which is what
we are primarily interested in -- does not (significantly) improve. As
a consequence, the $\Ng = 100$ case is assumed to be the standard grid
for the integral space as far as the dynamical modeling is concerned
to give an unbiased solution. Once the non linear parameters have been
recovered through the evidence maximization routine, it is possible to
start from the obtained results to make a more detailed and expensive
study of the distribution function using a higher number of TICs.

\subsection{Model uncertainties}
\label{error.analysis}

\begin{figure*}
\begin{center}
\resizebox{1.00\hsize}{!}{\includegraphics[angle=-90]{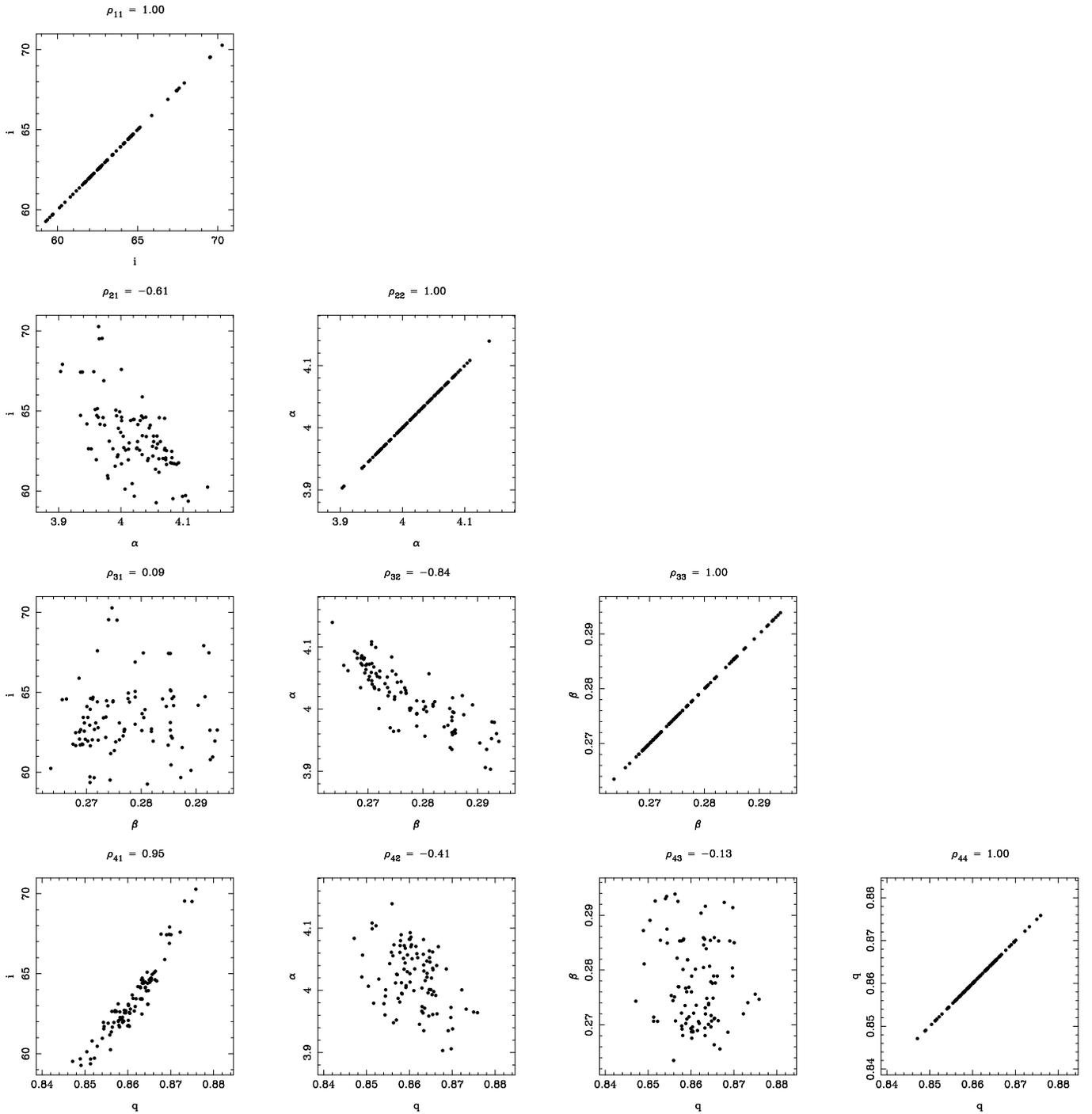}}
\caption{Graphical visualization of the lower triangle of the
(symmetric) correlation matrix for the parameters recovered from the
non linear reconstruction of $\Nrealiz$ random realizations of the
test data. In the shown panels the four non linear parameters $i$,
$\alpha$, $\beta$ and $q$ are plotted two by two against each other;
for each panel, the corresponding value $\rho_{ij}$ of the correlation
matrix is also indicated.}
\label{fig.corr.matrix}
\end{center}
\end{figure*}

To determine the scatter in the recovered parameters $\veceta$ we
performed the full non linear reconstruction on a set of $\Nrealiz$
random realizations of the test data. The statistical analysis of the
results is summarized in Table~\ref{table.error}. On examining the
$95\%$ confidence intervals, the parameters appear to be quite tightly
constrained, with the partial exception of the inclination angle which
spans an interval of approximately $10^{\circ}$. The means (and the
medians) of the parameters $\alpha$, $\beta$ and $q$ are very close to
the ``true values'' of the reference model, while the mean of
parameter $i$ is slightly more skewed. The two sets of parameters
($i$, $q$) and ($\alpha$, $\beta$) are clearly significantly
correlated when plotted against each other (see
Fig.~\ref{fig.corr.matrix} for a graphical representation of the
correlation matrix). Both correlations can be understood. The former
is due to the fact that the projected lens potential is nearly
invariant when the inclination is increased and flattening is
decreased simultaneously; whereas, in the latter case the requirement
that the lens mass enclosed by the lensed images be very similar
between lens models causes the lens strength and density slopes to vary
in concordance.

Although the starting parameters of model $\model_{0}$ are
very different from the ``true'' solution given by $\model_{\rm ref}$,
in all the considered cases the recovered parameters end up close to
the $\model_{\rm ref}$ ones, and there is no case in which the
solution remains anchored in a really far-off local minimum.  Hence,
even in the case of our chosen lens potential, which has global
degeneracies in the lensing observables (see Section~\ref{biases}),
these degeneacies are broken through the inclusion of stellar
kinematic data.  This indeed shows that the combination of lensing and
stellar kinematic information is a very promising tool to break
degeneracies of the galaxy mass models.

\begin{table}
\begin{center}
\caption{Summary of the results of the non linear reconstruction
algorithm applied on a set of $\Nrealiz$ random realizations of the
test data.}
\smallskip
\begin{tabular}{ c c c c c c c c }
\hline
\noalign{\smallskip}
  & & median & mean & $95\%$ C.I. & $\model_{\rm ref}$ \\
\noalign{\smallskip}
\hline
\noalign{\smallskip}
           & $i\;\textrm{[deg]}$ & 63.0   & 63.3  & [59.5,  69.5]  & 60.0  \\
non linear & $\talp$             & 4.02   & 4.02  & [3.94,  4.10]  & 4.05  \\
parameters & $\beta$             & 0.276  & 0.277 & [0.266, 0.293] & 0.280 \\
           & $q$                 & 0.861  & 0.861 & [0.849, 0.873] & 0.850 \\
\noalign{\smallskip}
\hline
\end{tabular}
\label{table.error}
\end{center}
\end{table}

\section{Conclusions and future work}
\label{conc}

We have presented and implemented a complete framework to perform a
detailed analysis of the gravitational potential of (elliptical) lens
galaxies by combining, for the first time, in a fully self-consistent
way both gravitational lensing and stellar dynamics information.

This method, embedded in a Bayesian statistical framework, enables one
to break to a large extent the well-known degeneracies that constitute
a severe hindrance to the study of the early-type galaxies (in
particular those at large distances, i.e.\ $z \gtrsim 0.1$) when the
two diagnostic tools of gravitational lensing and stellar dynamics are
used separately.  By overcoming these difficulties, the presented
methodology provides a new instrument to tackle the major
astrophysical issue of understanding the formation and evolution of
early-type galaxies.

The framework is very general in its scope and in principle can
accommodate an arbitrary (e.g.\ triaxial) potential $\Phi(\vecx)$. In
fact, if a combined set of lensing data (i.e.\ surface brightness
distribution of the lensed images) and kinematic data (i.e.\ surface
brightness distribution and velocity moments maps of the galaxy) is
provided for an elliptical lens galaxy, it is always possible, making
use of the same potential, to formulate the two problems of
lensed-image reconstruction and dynamical modeling as sets of coupled
linear equations to which the linear and non linear optimization
techniques described in Sect.~\ref{outline} and~\ref{bayes} can be
directly applied. 

More specifically, in the case of gravitational lensing the
non-parametric source reconstruction method (as illustrated in
Sect.~\ref{len}) straightforwardly applies to the general case of any
potential $\Phi(\vecx)$. In the case of dynamical modeling, a full
Schwarzschild method with orbital integration would be required; this
would constitute, however, only a mere technical complication which
does not modify the overall conceptual structure of the method.

\subsection{The \dynlen\ algorithm}

In practical applications, however, technical difficulties and
computational constraints must also be taken into account. This has
motivated the development, from the general framework, of the specific
working implementation described in this paper, which restricts itself
to axisymmetric potentials $\Phi(R, z)$ and two-integral stellar
phase-space distribution functions. This choice is an excellent
compromise between efficiency and generality.\footnote{This is
particularly true also in consideration of the currently available
data quality for distant early-type galaxies, for which the
information about the projected velocity moments is in general not
very detailed, and could not reliably constrain a sophisticated
dynamical model.} On one hand it allows one to study models which go
far beyond the simple spherical Jeans-modeling case, and on the other
hand, it has the invaluable advantage of permitting a dynamical
modeling by means of the two-integrals Schwarzschild method of
\citet{Vero.02}. This method (see Sect.~\ref{dyn}) is based on the
superposition of elementary building blocks (i.e.\ the TICs) directly
obtained from the two-integral distribution function, which
do not require computationally expensive orbit integrations.

More specifically, we have sped up this method by several orders of
magnitude by designing a fully novel Monte Carlo implementation (see
Appendix~\ref{MCMC}). Hence, we are now able to construct a realistic
two-integral dynamical model and its observables (surface brightness
and line-of-sight projected velocity moments) in a time of the order
of $5 - 15$ s on a typical {\machine} machine.

The Bayesian approach (e.g. Sect.~\ref{bayes}) constitutes a
fundamental aspect of the whole framework. On a first level, the
maximization of the posterior probability, by means of linear
optimization, provides the most probable solution for a given data set
and an assigned model (which will be, in general, a function of some
non linear parameters $\veceta$, e.g.\ the potential parameters, the
inclination, the position angle) in a fast and efficient way. A
solution (i.e.\ the source surface brightness distribution for lensing
and the distribution function for dynamics), however, can be obtained
for \emph{any} assigned model. The really important and challenging
issue is instead the model comparison, that is, to objectively
determine which is the ``best'' model for the given data set or, in
other words, which is the ``best'' set of non linear parameters
$\veceta$. Bayesian statistics provides the tool to answer these
questions in the form of a merit function, the ``evidence'', which
naturally and automatically embodies the principle of Occam's
razor, penalizing not only mismatching models but also models which
correctly predict the data but are unnecessarily complex
\citep[e.g.][]{MacKay.92, MacKay.99, MacKay.03}. The problem of model
comparison thus becomes a non linear optimization problem (i.e.\
maximizing the evidence), for which several techniques are available
(see Sect.~\ref{nlin.opt}).

As reported in Sect.~\ref{test}, we have conducted successful tests of
the method, demonstrating that both the linear reconstruction and the
non linear optimization algorithms work reliably. It has been shown
that it is possible to recover within a few percent the values of the
non linear parameters of the reference model (i.e. the ``true'' model
used to generate the simulated data set), even when starting the
reconstruction from a very skewed and implausible (in terms of the
evidence value) initial guess for the non linear parameters. Such an
accurate reconstruction is a direct consequence of having taken into
account, beyond the information coming from gravitational lensing, the
constraints from stellar dynamics. Indeed, when the algorithm is run
considering {\sl only} the lensing data, degenerate solutions with
comparable or almost coincident values for the evidence are
found,\footnote{Because of the presence of noise in the data,
numerical errors and model inaccuracies, solutions for the non linear
parameters which differ from the parameters of the reference models
can be slightly favored by the evidence.} making it effectively
impossible to distinguish between these models. The crucial importance
of the information from dynamics is exhibited by the fact that, when
it is included in the analysis, the degeneracies are fully broken and
a solution very close to the true one is unambiguously recovered (see
Fig.~\ref{fig.degen} for an example).

Bearing in mind these limitations and their consequences, however, it
should also be noted that the full modularity of the presented
algorithm makes it fit to be used also in those situations in which
either the lensing or the kinematic observables are not
available. This would allow one, for example, to restrict the plausible
models to a very small subset of the full space of non linear
parameters, although a single non-degenerate ``best solution'' would
probably be hard or impossible to find.

\subsection{Future work}

Eventhough the methodology that we introduced in this paper works very
well and is quite flexible, we can foresee a number of improvements
for the near and far future, in order of perceived complexity. (i)
Exploration of the errors on the non linear parameters $\vec \eta$
through a MCMC method, even though this requires an extension of the
MCMC framework in the context of the Bayesian evidence, since one also
needs to explore variations in the lens parameters due to changes in
the hyperparameters, and not only changes in the posterior for fixed
hyperparameters. (ii) Implementation of additional potential (or
density) models or even a non-parametric or multi-pole expansion
description of the gravitational potential in the $(R,z$)-plane for
axisymmetric models. This allows more freedom for the galaxy potential
description. (iii) Implementing an approximate three-integral method
in axisymmetric potentials \citep[e.g.][]{Dehnen.93}. (iv) Including
an additional iterative loop around the posterior probability
optimization, one can construct stellar phase-space distribution
functions that are self-consistent, i.e.\ they generate the potential
for which they are solutions to the collisionless Boltzmann
equation. This would allow the stellar and dark-matter potential
contributions to be separated, a feature not yet part of the current
code. (v) A full implementation of Schwarzschild's method for
arbitrary potentials through orbital integration.

Besides these technical improvements, which are all beyond the scope
of this methodological paper, we also plan, in a future publication, a
set of additional performance tests to see to what level each of the
degeneracies (e.g. the mass sheet and mass anisotropy) in lensing and
stellar dynamics are broken and how far the simpler lensing plus
spherical Jeans approach \citep[e.g.][]{Treu.04, Koop.06} gives
(un)biased results. Such studies would allow us to better interpret
the results obtained in cases where spatially resolved stellar
kinematics is not available \citep[e.g.\ for faint very high redshift
systems;][]{Koop.02}.

As for the application, the algorithm described in this paper will be
applied in a full and rigorous analysis of the SLACS sample of massive
early-type lens galaxies \citep[see][]{Bolton.06, Treu.06, Koop.06}
for which the available data include \emph{Hubble Space Telescope}
Advanced Camera for Survey and NICMOS images of the galaxy surface
brightness distribution and lensed image structure and maps of the
first and second line-of-sight projected velocity moments (obtained
with VLT-VIMOS two-dimensional integral field spectroscopy, as part of
the Large Program and as a series of Keck long-slit spectra).

\begin{acknowledgements}

We are grateful to the anonymous referee for the insightful
comments. We acknowledge the Kavli Institute for Theoretical Physics
at the University of California, Santa Barbara for the warm
hospitality and lively scientific environment provided over the course
of the \emph{Applications of Gravitational Lensing} workshop. We are
grateful to Tommaso Treu, Chris Fassnacht, Giuseppe Bertin, Luca
Ciotti, Phil Marshall, Sherry Suyu, and Claudio Grillo for fruitful
discussions, as well as to the other SLACS team members, Adam Bolton,
Scott Burles, and Lexi Moustakas. M.~B. acknowledges the support from
an NWO program subsidy (project 614.000.417). L.~V.~E.~K. is supported
in part through an NWO-VIDI program subsidy (project 639.042.505). We
also acknowledge the continuing support of the European Community's
Sixth Framework Marie Curie Research Training Network Programme,
contract MRTN-CT-2004-505183 "ANGLES".

\end{acknowledgements}



\clearpage
\newpage

\appendix

\section{Regularization}
\label{reg}

We make use of a curvature regularization. This form of regularization
tries to put the second derivative between a pixel and the two
adjacent ones to zero and has been shown by \citet{Suyu.06} to be
optimal for the reconstruction of smooth distributions. The curvature
regularization has been chosen since we do not expect, in the majority
of cases, to have sharp intensity variations in the surface brightness
distribution of an extended source (for lensing) or in the
distribution function of a galaxy. However, other choices
of regularization can easily be implemented and ranked according
to their evidence \citep{Suyu.06}.

Following the notation of Sect.~\ref{bayes}, we indicate as $\vecx$
the linear parameter vector (or, more simply, the source) and as
$\matH$ the regularization matrix. Since the source is defined on a
rectangular grid of $N = N_{\rm row} \times N_{\rm col}$ elements,
$\matH$ actually consists of two matrices, $\matH_{\rm row}$ and
$\matH_{\rm col}$, which regularize the grid pixels along the rows and
the columns, respectively.
The horizontal regularization operator $\matH_{\rm row}$ is a square
matrix of rank $N$. In each row~$i$ the only non zero elements are
$h_{i, i-1} = +1, h_{i,i} = -2, h_{i, i+1} = +1$; the only exceptions
are the rows $1+k N_{\rm col}$ and $N_{\rm col} + k N_{\rm col}$ (with
$k = 0, 1, \dots, N_{\rm row} - 1$), where a zeroth-order
regularization is performed (i.e. $h_{i,i} = 1$ is the only non zero
term) to prevent the connection of pixels belonging to different rows
which are therefore physically uncorrelated.
Similarly, the vertical regularization operator $\matH_{\rm col}$ is
constructed such that in each row~$i$ all the elements are zero with
the exclusion of $h_{i, i-N_{\rm col}} = +1, h_{i,i} = -2, h_{i,
i+N_{\rm col}} = +1$; a zeroth-order regularization is applied for the
rows $1 \dots N_{\rm col}$ and $N - N_{\rm col} + 1 \dots N$.

\section{Normalization: setting the scale of lensing}
\label{defl.angle}

For the assigned three-dimensional gravitational potential $\Phi$, the
reduced deflection angle $\vecalp$ is given by \citep*[e.g.][]{Schn.book}
\begin{equation}
\label{alpha.dim}
\vecalp(x', y') = \frac{2}{c^{2}} \frac{\Dds}{\Ds}
                  \int_{-\infty}^{+\infty} \nabla_{\vecxi} 
                  \Phi(\vecxi, z') \, \de z' ,
\end{equation}
where $z'$ is the line-of-sight coordinate, $(x', y') \equiv \vecxi$
are the sky coordinates and $\nabla_{\vecxi}$ denotes the
two-dimensional gradient operator in the plane of the sky; $c$ is the
speed of light expressed in the same units as the value of ${|\Phi|}^{1/2}$.
If the gradient operator, which does not depend on $z'$, is taken out
of the integral and the potential is conveniently written as
\begin{equation}
\label{Phi.adim}
\Phi(\vecxi, z') = \Phiz \times \tPhi (\txi, \tz'),
\end{equation}
where $\Phiz$ is the normalization constant in the most suitable
physical unit (in our case km s$^{-1}$) and $\tPhi$ is a function of
the dimensionless coordinates expressed as angles in arcseconds ($\txi
\equiv \frac{648000}{\pi} \vecxi/\Dd$, $\tz' \equiv \frac{648000}{\pi}
z'/\Dd$), then the deflection angle assumes the expression
\begin{equation}
\label{alpha.adim}
\vecalp(\tx, \ty) = \talp \, \nabla_{\txi} \left[
                    \int_{-\infty}^{+\infty} \tPhi (\txi, \tz') 
                    \, \de z' \right],
\end{equation}
where 
\begin{equation}
\label{lens.str}
\talp \equiv \frac{6.48 \times 10^{5}}{\pi} \frac{2 \Phiz}{c^{2}} 
             \frac{\Dds}{\Ds}
\end{equation}
is the lens strength in arcseconds. The parameter $\talp$ sets the
scale for the lensing and, therefore, is always included in the
parameter vector $\veceta$. Eq.~(\ref{lens.str}) openly displays how
intimately the lens strength is connected to the normalization of the
three-dimensional potential (the same used for the dynamical modeling)
within our joint method.

\section{A MCMC implementation of the two-integral 
axisymmetric Schwarzschild method}
\label{MCMC}

In this Appendix we describe the numerical implementation of the
two-integral axisymmetric Schwarzschild method of \citet{Vero.02} that
we developed to significantly accelerate the construction of
the dynamical model, i.e.\ the projected and PSF-convoluted model
observables generated by the TIC library.

As a first step, we construct a library composed of $\Ng = \nE \times
\nLz$ TICs in the given potential $\Phi$ (here $\nLz$ is an even
number). We consider a grid (linear or logarithmic) in the circular
radius $\Rc$ with $\nE$ samplings between $R_{\rm c, min}$ and $R_{\rm
c, max}$. The range is chosen to include most of the mass or, if the
mass is infinite for the potential $\Phi$, to provide a satisfactory
sampling of the potential profile in the radial direction. For each
$\Rc$ the circular speed $\vc$ is calculated as
\begin{equation}
\label{vc}
\vc^{2}(\Rc) = \Rc \left. \frac{\partial \Phi}{\partial R} 
               \right|_{(\Rc, 0)}
\end{equation}
and the angular momentum of the circular orbit $L_{z, {\rm max}} = \Rc
\vc$ is set. Computing the energy $E_{\rm c} \equiv E(\Rc) = V_{\rm
eff}(\Rc, 0)$ of the circular orbit at $\Rc$, the radial grid is
immediately translated into a sampling in energy. For each value of
$E_{\rm c}$, the grid in the normalized angular momentum $\eta \equiv
L_{z}/L_{z, {\rm max}}$ is constructed by sampling (linearly or
logarithmically) $\nLz/2$ values between the minimum $\eta_{\rm min} =
0$ and the maximum $\eta_{\rm max} = 1$. (For numerical reasons, the
grid is actually not sampled between these extrema, but between
$\eta_{\rm min} = \epsilon$ and $\eta_{\rm max} = 1 - \epsilon$, with
$\epsilon \ll 1$). To take the odd part of the distribution function
into account as well, we likewise consider the $\nLz/2$ negative
values for the angular momentum, $\eta = -1 \dots \eta = 0$, on a
mirror grid.

We also need to define a suitable coordinate frame for the
galaxy. Since the system is axisymmetric, we adopt the cylindrical
coordinates $(R,\varphi,z)$, with the origin on the center of the
galaxy. If the galaxy is observed at an inclination~$i$ and $\PA$ is
the position angle (defined as the angle measured counterclockwise
between the north direction and the projected major axis of the
galaxy), the projected coordinates $(x', y', z')$ are given by
\begin{equation}
\label{coord}
\left\{
\begin{array}{l}
x' = R (\cos \PA \sin \varphi + \sin \PA \cos i \cos \varphi) -
     z \sin \PA \sin i \\
 \\
y' = R (\sin \PA \sin \varphi - \cos \PA \cos i \cos \varphi) +
     z \cos \PA \sin i \\
 \\
z' = R \sin i \cos \varphi + z \cos i .
\end{array}
\right.
\end{equation}
Here $z'$ is measured along the line of sight, while $x'$ and $y'$ are
in the plane of sky and are directed (respectively) along the
projected major and minor axes of the galaxy.

For any TIC, we populate the surface inside the zero velocity curve
with $\nTIC$ particles of mass (or equivalently luminosity) $m$ by
means of a Markov-Chain Monte Carlo routine whose probability
distribution is given by Eq.~(\ref{TIC.rho}) for the density. This
effectively corresponds to numerically reproducing the density
$\rho(R,z)$, fixing at the same time the total mass $\mTIC = m \nTIC$
for each TIC. However, since $\rho(R,z) \propto 1/R$, the surface
density of the torus ``wrapped'' onto the meridional plane (denoted as
$\varsigma$) is constant, i.e.,
\begin{equation}
\label{TIC.wrap}
\varsigma_{j}(R,z) \equiv \int_{0}^{2\pi} \rho_{j}(R,z) R \de \varphi = 
\left\{
\begin{array}{ll}
2 \pi^{2} \Cj & \textrm{inside ZVC} \\
& \\
0 & \textrm{elsewhere}. \\
\end{array}
\right.
\end{equation}
One can take advantage of this property to greatly simplify the
Markov-Chain Monte Carlo routine. Now it is only necessary to
uniformly populate the meridional plane. For each particle a pair of
coordinates $(R,z)$ (within some interval which encompasses the ZVC)
is randomly generated. If it falls outside the ZVC, the particle is
``rejected'' and another one is generated. If the coordinates are
located inside the ZVC, a random value in the interval $[0, 2\pi)$ is
chosen for the azimuthal coordinate in order to have a complete tern
$(R, \varphi, z)$, and the particle counts toward the total of $\nTIC$
drawings. This procedure yields at the same time the surface $\Azvc$
enclosed by the ZVC in the meridional plane (effectively obtained via
Monte Carlo integration), which is required for the normalization of
the $\gamma_{j}$ (see Appendix~\ref{normalization}). 

\begin{figure}
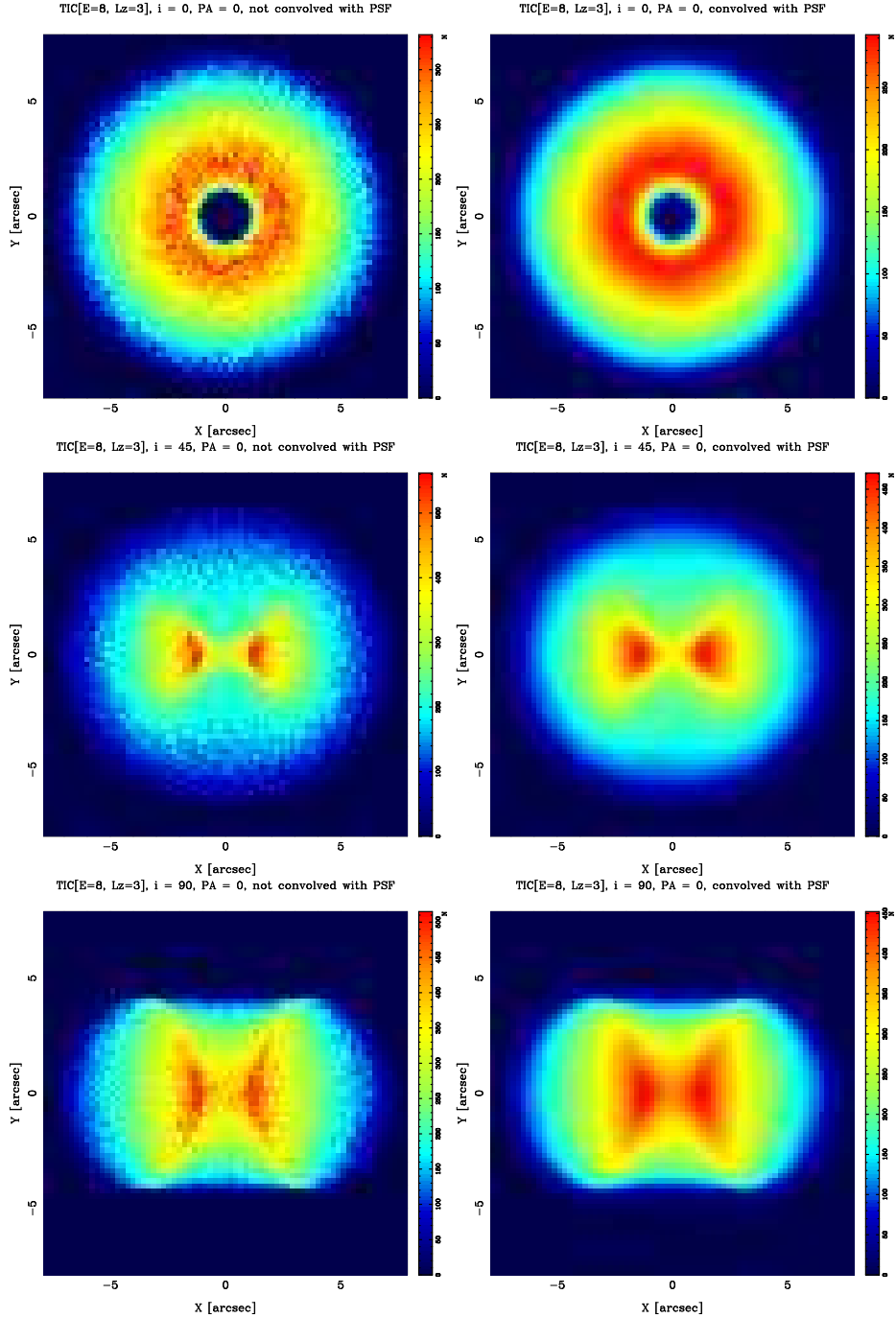
 
\begin{center}

\centerline{
  \mbox{\includegraphics[angle=-90,scale=.31]{TIC.00.ps}}
  \mbox{\includegraphics[angle=-90,scale=.31]{TIC.00.PSF.ps}}
}
\vspace{0.0cm}

\centerline{
  \mbox{\includegraphics[angle=-90,scale=.31]{TIC.45.ps}}
  \mbox{\includegraphics[angle=-90,scale=.31]{TIC.45.PSF.ps}}
}
\vspace{0.0cm}

\centerline{
  \mbox{\includegraphics[angle=-90,scale=.31]{TIC.90.ps}}
  \mbox{\includegraphics[angle=-90,scale=.31]{TIC.90.PSF.ps}}
}

\caption{As an illustration of the method described in the text, we
show for a given TIC (obtained with $\nTIC = 5 \times 10^{5}$
particles) the density distribution projected, from top to bottom, at
$i = 0\degr$ (face-on), $i = 45\degr$, and $i = 90\degr$ (edge-on). In
the left column the density distribution is not convolved with the
PSF; in the right column it is convolved with a Gaussian PSF of standard
deviation $\sigma_{x} = \sigma_{y} = 0.10\arcsec$.}
\label{fig.TIC}
\end{center}
\end{figure}

With this method the computation of all the projected quantities is
fast and straightforward. For each ``accepted'' mass point we know the
cylindrical coordinates $(R, \varphi, z)$; associated with it are also
the velocity moments defined by Eqs.~(\ref{TIC.vphi})-(\ref{TIC.vz}).
Using the first two equations of the transformation from
equation~(\ref{coord}), the projected coordinates $(x', y')$ are
directly calculated. Casting the points on a grid on the sky plane and
summing up all the points in the same pixel then reproduces
numerically the projected surface brightness distribution $\SB_{j}$
(see Figure~\ref{fig.TIC} for an illustration). The line-of-sight
velocity moments associated with each point in the sky plane (but
possibly on a different grid) are obtained in an analogous way (but
making use now of the third equation of [\ref{coord}]) from the
corresponding unprojected quantities
\begin{equation}
\label{vz}
\vz = - \vphi \sin i \sin \varphi ,
\end{equation}
\begin{equation}
\label{vqz}
\vqz = \left( \vqR \cos^{2} \varphi + \vqphi \sin^{2} \varphi \right) 
       \sin^{2} i + \vqR \cos^{2} i.
\end{equation}
In analogy with the surface brightness, the first and second
line-of-sight moments associated with each mass point inside a given
pixel are summed up. This gives the quantities $\Sigma \vz$ and
$\Sigma \vqz$.

The effect of the PSF is taken into account by simply convolving the
projected surface brightness or weighted velocity moments calculated
on their respective grids (preferably oversampled) with the PSF profile
sampled on the same grid. This operation can be numerically performed
in a very efficient way through several FFTs (fast Fourier transforms).

This numerical implementation is dramatically faster than the semi
analytic approach (at the expense of some numerical noise). On a
machine with a {\machine} processor, the whole process of calculating
the projected quantities $\SB$, $\vz$, and $\vqz$ convolved with the
PSF for 1400 TICs takes about 3 minutes (with $\nTIC = 10^{5}$). This
figure should be compared with the $\sim 30$ minutes required by
\citet{Vero.02} to calculate (on a 1 GHz machine) {\sl only} the
projected density without PSF convolution for an equal number of TICs.

\section{Normalization of the TICs weights}
\label{normalization}

In this Appendix we illustrate how the reconstructed adimensional
weights $\gamma_{j}$ are translated into the dimensional distribution
function values $\textrm{DF}(E_{j}, L_{z,j})$.
For a two-integral distribution function, the density is given by the
formula \citep[e.g.][]{Binn.book}
\begin{equation}
\label{rho-DF}
\varsigma(R,z) = 2 \pi R \rho(R,z) = 
                 4 \pi^{2} \int_{0}^{V} \de E \int_{L_{z}^{2} < 2 (V - E)R^{2}}
                 \textrm{DF}(E, L_{z}) \de L_{z},
\end{equation}
where $\varsigma (R,z)$ is the surface density ``wrapped'' in the
meridional plane (cf. definition~[\ref{TIC.wrap}]). If we assume that
$\textrm{DF}$ can be considered approximately constant over the cell
$\de E_{j} \de L_{z,j}$ of area $\AELz$ in the integral space, and we
remember the properties of the TICs, then we have
\begin{equation}
\label{sig-DF}
4 \pi^{2} \textrm{DF}(E_{j}, L_{z,j}) \de E_{j} \de L_{z,j} = 
\gamma_{j} \varsigma_{j}(R,z) .
\end{equation}

In the previous formula, $\varsigma_{j}$ is the ``wrapped'' surface
density generated by $\textrm{TIC}_{j}$, specified by the pair
$(E_{j}, L_{z,j})$, which is constant inside the ZVC and zero
elsewhere (see Eq.~[\ref{TIC.wrap}]):
\begin{equation}
\label{Cj}
\varsigma_{j} = 2 \pi^{2} \Cj = \frac{\mTIC}{\Azvc} \qquad
\textrm{(inside the ZVC)} ;
\end{equation}
here $\Azvc$ denotes the area enclosed by the ZVC in the meridional
plane (which can be calculated as described in Appendix~\ref{MCMC})
and $\mTIC = m \nTIC$ is the fixed TIC mass (all the TICs have equal
mass by construction).

Combining Eqs.~(\ref{sig-DF}) and~(\ref{Cj}) we find the desired
relation 
\begin{equation}
\label{gamma-DF}
\textrm{DF} (E_{j}, L_{z,j}) = 
            \gamma_{j} \frac{m \nTIC}
            {4 \pi^{2} \Azvc \AELz} ,
\end{equation}
which translates the weights $\gamma_{j}$ into distribution function
values expressed in the standard physical units of mass length$^{-3}$
velocity$^{-3}$. If $m$, in the numerator of the right-hand side of
Eq.~(\ref{gamma-DF}), is omitted or divided by a mass-to-light ratio
coefficient $\Gamma$, the resulting distribution function is expressed
in terms of (respectively) number or luminosity phase-space density.

\section{The evidence formula}
\label{evidence}

In this Appendix we use the same notation as Sect.~\ref{bayes}, and we
indicate as $\Nx$ and $\Nb$, respectively, the number of elements in
the linear parameter vector $\vecx$ and in the data vector $\vecb$.
If the assumptions made in Sect.~\ref{bayes}, viz., Gaussian noise and
quadratic functional form of the regularization term $\Es(\vecx)$ with
minimum in $x_{\rm reg} = \vec{0}$, are valid, then the logarithm of
the evidence has the expression \citep[e.g.][]{Suyu.06}

\begin{eqnarray} 
\label{evid}
\log P(\vecb | \lambda, \matA, \matH) & = & 
    - \frac{1}{2} {(\matA \vecx - \vecb)}^{\rm T} \Cv (\matA \vecx - \vecb) 
    - \frac{\lambda}{2} {|| \matH \vecx ||}^{2} 
    - \frac{1}{2} \log \left[ \det \left( 
      \matAt \Cv \matA + \lambda \matHt \matH \right) \right] 
      \nonumber \\
& & + \frac{\Nx}{2} \log \lambda
    + \frac{1}{2} \log \left[ \det \left( \matHt \matH \right) \right] 
    - \frac{\Nb}{2} \log (2 \pi)
    + \frac{1}{2} \log \left( \det \Cv \right).
\end{eqnarray}

The expression of the evidence in the case of lensing and dynamics is
immediately obtained by rewriting Eq.~(\ref{evid}) with the notation
of Sections~\ref{len} and~\ref{dyn}.

\section{The deflection angle for Evans' power-law galaxies}
\label{evans}

All the relevant quantities for the Evans' power-law galaxy models
which are used in Sect.~\ref{test} are analytic (refer to
\citeauthor{Evans.94a} \citeyear{Evans.94a} and \citeauthor{Evans.94b}
\citeyear{Evans.94b} for the full expressions). The lensing deflection
angle $\vecalp$ can be calculated from the potential from
Eq.~(\ref{evans.pot}), resulting in:
\begin{equation}
\label{evans.alpha}
\left\{
\begin{array}{ccl}
\alpha_{x'} & = & \displaystyle \frac{2\sqrt{\pi}}{c^{2}} \frac{\Dds}{\Dd}
		  \frac{\Gamma \left( \frac{\beta+1}{2} \right)}
		       {\Gamma \left( \frac{\beta+2}{2} \right)}
                  \beta \Rcore^{\beta} \Phiz \frac{q}{q'} 
		  \frac{x'}{\left( \Rcore^{2} + {x'}^{2} + 
		  {y'}^{2}/{q'}^{2} \right)^{\frac{\beta+1}{2}}}\\
& & \\
\alpha_{y'} & = & \displaystyle \frac{2\sqrt{\pi}}{c^{2}} \frac{\Dds}{\Dd}
		  \frac{\Gamma \left( \frac{\beta+1}{2} \right)}
		       {\Gamma \left( \frac{\beta+2}{2} \right)}
                  \beta \Rcore^{\beta} \Phiz \frac{q}{q'} 
		  \frac{y'/{q'}^{2}}{\left( \Rcore^{2} + {x'}^{2} + 
		  {y'}^{2}/{q'}^{2} \right)^{\frac{\beta+1}{2}}}\\
\end{array}
\right.
\end{equation}
where $(x', y')$ are the coordinates in the sky plane (which is defined
as the plane orthogonal to the line of sight $z'$), $q' =
\sqrt{\cos^{2} i + q^{2} \sin^{2} i}$ is the projected axis ratio, and
$\Gamma$ is the gamma function.


\begin{thebibliography}{}

\bibitem[Arnaboldi et al.(1996)]{Arna.96} 
Arnaboldi, M., et al.\ 1996, \apj, 472, 145

\bibitem[Barnes(1992)]{Barnes.92} 
Barnes, J.~E.\ 1992, \apj, 393, 484 

\bibitem[Bender et al.(1993)Bender, Burstein, \& Faber]{Bend.93}
Bender, R., Burstein, D., \& Faber, S.M. 1993, \apj, 411, 153

\bibitem[Bernardi et al.(2003)]{Bern.03} 
Bernardi, M., et al. 2003, \aj, 125, 1882

\bibitem[Bertin et al.(1994)]{Bert.94} 
Bertin, G., et al.\ 1994, \aap, 292, 381

\bibitem[Binney(1981)]{Binn.81}
Binney, J.\ 1981, \mnras, 196, 455

\bibitem[Binney \& Tremaine(1987)]{Binn.book}
Binney, J. \& Tremaine, S.\ 1987, \emph{Galactic Dynamics} (Princeton:
Princeton Univ. Press)

\bibitem[Bolton et al.(2006)]{Bolton.06}
Bolton, A.~S., Burles, S., Koopmans, L.~V.~E., Treu, T., \& Moustakas, L.~A.\ 
2006, \apj, 638, 703 

\bibitem[Borriello et al.(2003)Borriello, Salucci, \& Danese]{Borr.03} 
Borriello, A., Salucci, P., \& Danese, L.\ 2003, \mnras, 341, 1109

\bibitem[Bower et al.(1992)Bower, Lucey, \& Ellis]{Bower.92} 
Bower, R.G., Lucey, J.R., \& Ellis, R.S. 1992, \mnras, 254, 589

\bibitem[Brewer \& Lewis(2006a)]{Brewer.06a} Brewer, B.~J., \& 
Lewis, G.~F.\ 2006, \apj, 637, 608 

\bibitem[Brewer \& Lewis(2006b)]{Brewer.06b} Brewer, B.~J., \& 
Lewis, G.~F.\ 2006, \apj, 651, 8 

\bibitem[Byrd et al.(1995)Byrd, Lu, \& Nocedal]{Byrd.95}
Byrd, R.~H., Lu, P., \& Nocedal, J.\ 1995, SIAM Journal on Scientific
and Statistical Computing, 16, 5, 1190

\bibitem[Cappellari et al.(2006)]{Capp.06}
Cappellari, M., Bacon, R., Bureau, M., Damen, M.~C., Davies, R.~L., de
Zeeuw, P.~T., Emsellem, E., Falc\'on-Barroso, J., Krajnovi\'c, D.,
Kuntschner, H., McDermid, R.~M., Peletier, R.~F., Sarzi, M., van den
Bosch, R.~C.~E., van de Ven, G.\ 2006 \mnras, 366, 1126

\bibitem[Carollo et al.(1995)]{Caro.95} 
Carollo, C.~M., de Zeeuw, P.~T., van der Marel, R.~P., Danziger,
I.~J., \& Qian, E.~E.\ 1995, \apjl, 441, L25

\bibitem[Cohn et al.(2001)]{Cohn.01} Cohn, J.~D., Kochanek, 
C.~S., McLeod, B.~A., \& Keeton, C.~R.\ 2001, \apj, 554, 1216 

\bibitem[Cole et al.(2000)]{Cole.00}
Cole, S., Lacey, C.~G., Baugh, C.~M., \& Frenk, C.~S.\ 2000, \mnras,
319, 168

\bibitem[Cousins(1995)]{Cous.95}
Cousins, R.~D.\ 1995, American Journal of Physics, 63, 398

\bibitem[Cretton et al.(1999)]{Cret.99}
Cretton, N., de Zeeuw, P.~T., van der Marel, R.~P., \& Rix, H.-W.\ 1999,
\apjs, 124, 383

\bibitem[de Zeeuw et al.(2002)]{deZe.02} 
de Zeeuw, P.~T., Bureau, M., Emsellem, E., Bacon, R., Carollo, C.~M.,
Copin, Y., Davies, R.~L., Kuntschner, H., Miller, B.~W., Monnet, G.,
Peletier, R.~F., Verolme, E.~K.\ 2002, \mnras, 329, 513

\bibitem[Dehnen \& Gerhard(1993)]{Dehnen.93} Dehnen, W., \& 
Gerhard, O.~E.\ 1993, \mnras, 261, 311 

\bibitem[Djorgovski \& Davis(1987)]{Djor.87}
Djorgovski, S., \& Davis, M. 1987, \apj, 313, 59

\bibitem[Dobke \& King(2006)]{Dobke.06} Dobke, B.~M., \& King, 
L.~J.\ 2006, \aap, 460, 647 

\bibitem[Dressler et al.(1987)]{Dres.87} 
Dressler, A., Lynden-Bell, D., Burstein, D., Davies, R.~L., Faber,
S.M., Terlevich, R., \& Wegner, G. 1987, \apj, 313, 42

\bibitem[Dye \& Warren(2005)]{Dye.05} Dye, S., \& Warren, 
S.~J.\ 2005, \apj, 623, 31 

\bibitem[Evans(1994)]{Evans.94a}
Evans, N.~W.\ 1994, \mnras, 267, 333

\bibitem[Evans \& de Zeeuw(1994)]{Evans.94b}
Evans, N.~W. \& de Zeeuw, P.~T.\ 1994, \mnras, 271, 202

\bibitem[Evans \& Witt(2003)]{Evans.03} Evans, N.~W., \& Witt, 
H.~J.\ 2003, \mnras, 345, 1351 

\bibitem[Fabbiano(1989)]{Fabb.89} 
Fabbiano, G.\ 1989, \araa, 27, 87

\bibitem[Falco et al.(1985)Falco, Gorenstein, \& Shapiro]{Falco.85}
Falco, E.~E., Gorenstein, M.~V., \& Shapiro, I.~I. 1985, ApJ, 289, L1

\bibitem[Ferreras et al.(2005)]{Fer.05} Ferreras, I., Saha, 
P., \& Williams, L.~L.~R.\ 2005, \apjl, 623, L5 

\bibitem[Ferrarese \& Merritt(2000)]{Ferr.00} 
Ferrarese, L., \& Merritt D. 2000, \apj, 539, L9

\bibitem[Franx et al.(1994)Frank, van Gorkom, \& de Zeeuw]{Franx.94} 
Franx, M., van Gorkom, J.~H., \& de Zeeuw, T.\ 1994, \apj, 436, 642

\bibitem[Frenk et al.(1988)]{Frenk.88} 
Frenk, C.~S., White, S.~D.~M., Davis, M., \& Efstathiou, G.\ 1988,
\apj, 327, 507

\bibitem[Gavazzi et al.(2007)]{Gava.07} Gavazzi, R., Treu, T., Rhodes,
J.~D., Koopmans, L.~V.~E., Bolton, A.~S., Burles, S., Massey, R., \&
Moustakas, L.~A. 2007, ApJ, in press

\bibitem[Gebhardt et al.(2000)]{Gebh.00} 
Gebhardt, K., et al. 2000, \apj, 539, L13

\bibitem[Gerhard(1993)]{Gerh.93} Gerhard, O.~E.\ 1993, \mnras, 
265, 213 

\bibitem[Gerhard et al.(2001)]{Gerh.01} 
Gerhard, O., Kronawitter, A., Saglia, R.~P., \& Bender, R.\ 2001, \aj,
121, 1936

\bibitem[Gerhard(2006)]{Gerh.06} Gerhard, O.\ 2006, Planetary 
Nebulae Beyond the Milky Way, 299

\bibitem[Gray \& Kolda(2005)]{GrKo05} 
Gray, G.~A. \& Kolda, T.~G.\ 2006, ACM T. Math. Software, 32, 3, 485

\bibitem[Guzman et al.(1992)]{Guzm.92} 
Guzman, R., Lucey, J.R., Carter, D., \& Terlevich, R.J. 1992, \mnras,
257, 187

\bibitem[James(1994)]{minuit}James, J.\ 1994, MINUIT - Function
Minimization and Error Analysis, CERN Program Library entry D506 

\bibitem[Kochanek(1995)]{Koch.95} Kochanek, C.~S.\ 1995, \apj, 
445, 559 

\bibitem[Kolda(2005)]{Ko05} 
Kolda, T.~G.\ 2005, SIAM J. Optimization, 16, 2, 563

\bibitem[Koopmans \& Treu(2002)]{Koop.02}
Koopmans, L. V. E. \& Treu, T.\ 2002, \apj, 568, L5

\bibitem[Koopmans \& Treu(2003)]{Koop.03}
Koopmans, L. V. E. \& Treu, T.\ 2003, \apj, 583, 686

\bibitem[Koopmans(2004)]{Koop.04} Koopmans, L.V.E., Proceedings of Science,
published by SISSA; Conference: ``Baryons in Dark Matter Haloes'',
Novigrad, Croatia, 5-9 October 2004; editors: R.-J. Dettmar, U. Klein,
P. Salucci, 66

\bibitem[Koopmans(2005)]{Koop.05}
Koopmans, L.~V.~E.\ 2005, \mnras, 363, 1136

\bibitem[Koopmans et al.(2006)]{Koop.06}
Koopmans, L.~V.~E., Treu, T., Bolton, A.~S., Burles, S., \& Moustakas, L.~A.\ 
2006, \apj, 649, 599

\bibitem[Kormann et al.(1994)]{Korm.94} Kormann, R., Schneider, 
P., \& Bartelmann, M.\ 1994, \aap, 284, 285

\bibitem[Liddle et al.(2007)]{Liddle.07} Liddle, A.~R, Corasaniti,
P.~S., Kunz, M., Mukherjee, P., Parkinson, D., \& Trotta, R.\ 2007,
astro-ph/0703285v1

\bibitem[Loewenstein \& White(1999)]{Loew.99} 
Loewenstein, M., \& White, R.~E.\ 1999, \apj, 518, 50

\bibitem[Ma(2003)]{Ma.03} Ma, C.-P.\ 2003, \apjl, 584, L1 

\bibitem[MacKay(1992)]{MacKay.92}
MacKay, J. C.\ 1992, PhD thesis

\bibitem[MacKay(1999)]{MacKay.99}
MacKay, J. C.\ 1999, Neural Comp, 11, 1035

\bibitem[Mackay(2003)]{MacKay.03} Mackay, D.~J.~C.\ 2003,
\emph{Information Theory, Inference and Learning Algorithms}
(Cambridge: Cambridge University Press)

\bibitem[Magorrian et al.(1998)]{Mago.98} 
Magorrian J., et al. 1998, \aj, 115, 2285

\bibitem[Marshall(2006)]{Mars.06} Marshall, P.\ 2006, \mnras, 
372, 1289 

\bibitem[Matsushita et al.(1998)]{Mats.98} 
Matsushita, K., Makishima, K., Ikebe, Y., Rokutanda, E., Yamasaki, N.,
\& Ohashi, T.\ 1998, \apjl, 499, L13

\bibitem[Merritt(1985a)]{Merr.85a} Merritt, D. 1985a, \aj, 90, 1027

\bibitem[Merritt(1985b)]{Merr.85b} Merritt, D. 1985b, \mnras, 214, 25

\bibitem[Mould et al.(1990)]{Mould.90} 
Mould, J.~R., Oke, J.~B., de Zeeuw, P.~T., \& Nemec, J.~M.\ 1990, \aj,
99, 1823

\bibitem[Mu{\~n}oz et al.(2001)]{Munoz.01} Mu{\~n}oz, J.~A., 
Kochanek, C.~S., \& Keeton, C.~R.\ 2001, \apj, 558, 657 

\bibitem[Osipkov(1979)]{Osip.79} Osipkov L.~.P., 1979, Pis'ma
Astron. Zh., 5, 77

\bibitem[Pfenniger(1984)]{Pfen.84} 
Pfenniger, D.\ 1984, \aap, 141, 171

\bibitem[Press et al.(1992)]{NR.92}
Press, W.~H., Teukolsky, S.~A., Vetterling, W.~T., Flannery, B.~P.\ 1992,
\emph{Numerical Recipes in C} (Cambridge: Cambridge University Press)

\bibitem[Richstone(1980)]{Rich.80} 
Richstone, D.~O.\ 1980, \apj, 238, 103

\bibitem[Richstone(1984)]{Rich.84} 
Richstone, D.~O.\ 1984, \apj, 281, 100

\bibitem[Rix et al.(1997)]{Rix.97} 
Rix, H.-W., de Zeeuw, P.~T., Cretton, N., van der Marel, R.~P., \&
Carollo, C.~M.\ 1997, \apj, 488, 702

\bibitem[Romanowsky et al.(2003)]{Roma.03} 
Romanowsky, A.~J., Douglas, N.~G., Arnaboldi, M., Kuijken, K.,
Merrifield, M.~R., Napolitano, N.~R., Capaccioli, M., \& Freeman,
K.~C.\ 2003, Science, 301, 1696

\bibitem[Rusin \& Ma(2001)]{Rusin.01} Rusin, D., \& Ma, C.-P.\ 
2001, \apjl, 549, L33 

\bibitem[Rusin et al.(2002)]{Rusin.02} Rusin, D., Norbury, M., 
Biggs, A.~D., Marlow, D.~R., Jackson, N.~J., Browne, I.~W.~A., Wilkinson, 
P.~N., \& Myers, S.~T.\ 2002, \mnras, 330, 205 

\bibitem[Rusin et al.(2003)]{Rusin.03} Rusin, D., et al.\ 2003, 
\apj, 587, 143 

\bibitem[Rusin \& Kochanek(2005)]{Rusin.05} Rusin, D., \& 
Kochanek, C.~S.\ 2005, \apj, 623, 666 

\bibitem[Saglia et al.(1992)Saglia, Bertin, \& Stiavelli]{Sagl.92} 
Saglia, R.~P., Bertin, G., \& Stiavelli, M.\ 1992, \apj, 384, 433

\bibitem[Saha \& Williams(2006)]{Saha.06} Saha, P., \& 
Williams, L.~L.~R.\ 2006, \apj, 653, 936 

\bibitem[Schneider et al.(1992)Schneider, Ehlers, \& Falco]{Schn.book}
Schneider, P., Ehlers, J., \& Falco E.~E.\ 1992, \emph{Gravitational
Lenses} (Berlin: Springer Verlag)

\bibitem[Schneider et al. (2006)]{saasfee} Schneider, P., Kochanek,
C.~S., \& Wambsganss, J.\ 2006, Saas-Fee Advanced Course 33:
Gravitational Lensing: Strong, Weak and Micro

\bibitem[Schwarzschild(1979)]{Schw.79} 
Schwarzschild, M.\ 1979, \apj, 232, 236 

\bibitem[Seljak(2002)]{Selj.02} 
Seljak, U.\ 2002, \mnras, 334, 797

\bibitem[Suyu et al.(2006)]{Suyu.06} 
Suyu, S.~H., Marshall, P.~J., Hobson, M.~P., \& Blandford, R.~D.\ 2006,
\mnras, 371, 983

\bibitem[Tikhonov(1963)]{Tikh.63} 
Tikhonov, A.~N.\ 1963, Soviet Math. Dokl., 4, 1035

\bibitem[Toomre(1977)]{Toom.77} Toomre, A.\ 1977, in ``Evolution of
Galaxies and Stellar Populations'', Ed. B.M. Tinsley \& R.B. Larson
(New Haven: Yale University Observatory), 401

\bibitem[Treu \& Koopmans(2002)]{Treu.02}
Treu, T. \& Koopmans, L. V. E.\ 2002, \apj, 575, 87

\bibitem[Treu \& Koopmans(2003)]{Treu.03}
Treu, T. \& Koopmans, L. V. E.\ 2003, \mnras, 343, L29

\bibitem[Treu \& Koopmans(2004)]{Treu.04}
Treu, T. \& Koopmans, L. V. E.\ 2004, \apj, 611, 739

\bibitem[Treu et al.(2006)]{Treu.06}
Treu, T., Koopmans, L.~V.~E., Bolton, A.~S., Burles, S., \& Moustakas, L.~A.\ 
2006, \apj, 640, 662

\bibitem[Verolme \& de Zeeuw(2002)]{Vero.02}
Verolme, E.~K. \& de Zeeuw, P.~T.\ 2002, \mnras, 331, 959

\bibitem[Warren \& Dye(2003)]{Warr.03}
Warren, S.~J. \& Dye, S.\ 2003, \apj, 590, 673

\bibitem[Wayth et al.(2005)]{Wayth.05} Wayth, R.~B., Warren, 
S.~J., Lewis, G.~F., \& Hewett, P.~C.\ 2005, \mnras, 360, 1333 

\bibitem[Wayth \& Webster(2006)]{Wayth.06} Wayth, R.~B., \& 
Webster, R.~L.\ 2006, \mnras, 372, 1187 

\bibitem[White \& Frenk(1991)]{White.91} 
White, S.~D.~M., \& Frenk, C.~S.\ 1991, \apj, 379, 52

\bibitem[Winn et al.(2003)]{Winn.03} Winn, J.~N., Rusin, D., \& 
Kochanek, C.~S.\ 2003, \apj, 587, 80 

\bibitem[Wucknitz(2002)]{Wuck.02} 
Wucknitz, O.\ 2002, \mnras, 332, 951

\bibitem[Wucknitz et al.(2004)]{Wuck.04} Wucknitz, O., Biggs, 
A.~D., \& Browne, I.~W.~A.\ 2004, \mnras, 349, 14 

\bibitem[Zhu et al.(1997)Zhu, Byrd, \& Nocedal]{Zhu.97} Zhu, C., Byrd,
R.~H., \& Nocedal, J.\ 1997, ACM T. Math. Software, 23, 4, 550

\end{thebibliography}
\end{document}